\documentclass[fleqn,usenatbib]{mnras}

\usepackage{newtxtext,newtxmath}

\usepackage[T1]{fontenc}

\DeclareRobustCommand{\VAN}[3]{#2}
\let\VANthebibliography\thebibliography
\def\thebibliography{\DeclareRobustCommand{\VAN}[3]{##3}\VANthebibliography}

\usepackage{graphicx}	
\usepackage{amsmath}	
\usepackage{tikz}
\usetikzlibrary{positioning}
\usepackage{xcolor}

\newcommand{\myRed}[1]{\textcolor{black}{#1}}

\title[RTFAST-S: Emulation of X-ray reverberation]{RTFAST-Spectra: Emulation of X-ray reverberation mapping for active galactic nuclei}

\author[B. J. Ricketts et al.]{
Benjamin J. Ricketts$^{1,2}$\thanks{E-mail: b.ricketts@sron.nl},
Daniela Huppenkothen$^{1,2}$, Matteo Lucchini$^{1}$,
Adam Ingram$^{3}$,
\newauthor{Guglielmo Mastroserio$^{4}$, Matthew Ho$^{5}$ and Benjamin Wandelt$^{5,6}$} \\
$^{1}$Anton Pannekoek Institute, University of Amsterdam, Science Park 904, Amsterdam, 1098 XH, Netherlands\\
$^{2}$SRON, Niel Bohrweg 4, Leiden, 2033 CA, Netherlands\\
$^{3}$School of Mathematics, Statistics and Physics, Newcastle University, Herschel Building, Newcastle Upon Tyne, NE1 7RU, UK\\
$^{4}$University of Milan, Via Festa del Perdono, 7, 20122 Milano MI, Italy\\
$^{5}$ Sorbonne Université/CNRS, Institut d'Astrophysique de Paris, 98 bis Bd Arago, 75014 Paris, France\\
$^{6}$Department of Physics and Astronomy, Johns Hopkins University, 3400 N Charles Street, Baltimore, MD 21218, USA \\
}

\date{Accepted 2025 February 21. Received 2025 February 21; in original form 2024 December 13}

\pubyear{2024}

\begin{document}
\label{firstpage}
\pagerange{\pageref{firstpage}--\pageref{lastpage}}
\maketitle

\begin{abstract}
Bayesian analysis has begun to be more widely adopted in X-ray spectroscopy, but it has largely been constrained to relatively simple physical models due to limitations in X-ray modelling software and computation time. As a result, Bayesian analysis of numerical models with high physics complexity have remained out of reach. This is a challenge, for example when modelling the X-ray emission of accreting black hole X-ray binaries, where the slow model computations severely limit explorations of parameter space and may bias the inference of astrophysical parameters. Here, we present \texttt{RTFAST-Spectra}: a neural network emulator that acts as a drop in replacement for the spectral portion of the black hole X-ray reverberation model \texttt{RTDIST}. This is the first emulator for the \texttt{reltrans} model suite and the first emulator for a state-of-the-art x-ray reflection model incorporating relativistic effects with 17 physically meaningful model parameters. We use Principal Component Analysis to create a light-weight neural network that is able to preserve correlations between complex atomic lines and simple continuum, enabling consistent modelling of key parameters of scientific interest. We achieve a $\mathcal{O}(10^2)$ times speed up over the original model in the most conservative conditions with $\mathcal{O}(1\%)$ precision over all 17 free parameters in the original numerical model, taking full posterior fits from months to hours. We employ Markov Chain Monte Carlo sampling to show how we can better explore the posteriors of model parameters in simulated data and discuss the complexities in interpreting the model when fitting real data. 
\end{abstract}




\section{Introduction}
Compact objects remain one of the most intriguing phenomena in the known universe. They are part of some of the most energetic systems in the universe and they present unique challenges in a regime where our understanding of fundamental physics may break down. In particular, black holes are both interesting laboratory for fundamental physics and unique objects in their own right.

One way of probing compact objects is via the study of their accretion flows. At its most extreme, accretion enables us to measure their fundamental physical properties such as the mass and spin of black holes \citep{mcclintock2011measuring} and the Equation of State of neutron stars \citep{ozel2016masses}. Accreting black holes are also known to show energetic jets, and thus help us to better understand the physical conditions under which relativistic jets form \citep{nemmen2012universal}. Characterizing the interaction between these accretion processes with the compact object, and the geometry of the system are crucial for our understanding of the underlying extreme physics. 

Black hole accretion models have become significantly more complex as we come to better understand the phenomena that govern their behaviour and we gain access to increasingly varied data such as X-ray polarisation \citep{weisskopf2016imaging} and high cadence time series \citep{jansen2001xmm,gendreau2012neutron,uttley2014multi}. Many of the foremost challenges in black hole physics arise when we try to combine the rich information in these data sets with sophisticated theoretical models \citep{remillard2006x,sobolewska2009long}.

We are particularly interested in the inner-most regions of the accreting black hole, composed of an accretion disk (of which there are multiple interpretations of the thickness, truncation and magneto-hydrodynamics) that emits thermally \citep{shakura1973black}, a region of energetic electrons around the black hole (known as the corona) that emits in a non-thermal power law \citep{thorne1975cygnus,sunyaev1979hard}, and, under certain conditions, a jet which also emits non-thermally \citep{blandford1979relativistic,levinson1996jets,romero2017relativistic}. 
A portion of the photons emitted by the corona illuminates the accretion disk where it is reprocessed and generates what is called a reflection spectrum \citep{Fabian1989reflection}. Modelling these phenomena requires the computation of photon trajectories in a regime where relativistic effects dominate \citep{cunningham1973optical,dauser2013irradiation}, which in turn strongly depend on the black hole spin. The different parts of the inner regions all interact with each other to contribute to a complex total spectrum that also varies with time. By observing the correlated variability of the emitted radiation as a function of time and energy, we are able to measure the light travel time difference of the reflected photons compared to the non-thermal coronal photons. Proper modelling of this emission allows us to measure the size of the inner region, thus the mass of the black hole.

A fundamental challenge in modelling accreting sources is the high degree of parameter degeneracy: different parameter combinations may fit a data set similarly well, and therefore choices such as starting parameters, random seeds and optimization algorithm may determine the scientific conclusion. 
This is undesirable, as we require robust inferences of models and their parameters in order to advance our scientific understanding. 
One potential solution is to explore a wider range of parameter space using sampling algorithms like Markov Chain Monte Carlo (MCMC). This can be efficiently done in a Bayesian context, and allows us to report multi-dimensional probability distributions of the parameters given the data, rather than single-point best-fit values and errors derived using approximations. Posterior distributions are particularly useful when the parameter space is complex and multiple combinations of parameters may generate models that fit the data similarly well (also known as multi-modal probability distributions).

Some studies in X-ray spectral modelling have adopted Bayesian inference \citep{dias2024investigating,buchner2014x} but have to contend with significant computational burdens. Models that calculate physical parameters self-consistently can be computationally expensive, and the high number of model evaluations required by sampling algorithms can make inferring posteriors for real data slow. As these models are also often high-dimensional, the number of evaluations required to map the posterior space dramatically increases, creating an exponential increase in inference time. This leads to a trade-off between the complexity (and realism) of models and the computational feasibility of comparing these models to data. 

The numerical model \texttt{RTDIST} \citep[henceforth I22]{ingram2022} was developed to model the X-ray spectra and timing products from the corona and reflection spectrum of active galactic nuclei and X-ray binary black holes. \texttt{RTDIST} incorporates X-ray reflection and reverberation into one model that self-consistently calculates ionisation within the disk according to the geometry of the system. We further expand on the features of \texttt{RTDIST} and assumptions in section \ref{rtdist}. \texttt{RTDIST} currently takes approximately 0.5 seconds to evaluate the time averaged spectrum once on a single core of a typical laptop. Timing products such as the lags-energy spectrum take slightly longer. When using MCMC sampling to obtain Bayesian posteriors, this means that modelling a single data set can take weeks to months of computation time.

An alternative to directly computing the numerical model often applied in X-ray astronomy are interpolation table models. Many models are computed on a grid of parameters which will then be collated into a table. The table is used in linear interpolation to obtain model estimates for parameters not present in the table itself \citep{wilms2000absorption,garcia2013x}. This approach is much faster computationally, but is no longer feasible for high-dimensional models like \texttt{RTDIST} as the required number of stored, pre-computed models expand to the power of the number of model parameters, leading to unfeasibly large interpolation tables to be read into memory.

X-ray astronomy is not the first field of research, both within and outside of astronomy, to have faced this problem. Interpolation models can be (relatively) easily replaced by creating an emulator composed of flexible, parametric or non-parametric functions that replicate the relationship of input parameters and model outputs. Emulators have been used in a wide variety of scenarios within astronomy including: cosmological simulations \citep{heitmann2009coyote,kaushal2022necola,bird2019emulator}, cosmological matter power spectra \citep{spurio2022cosmopower,heitmann2013coyote}, optical and UV energy spectra of galaxies and Type Ia supernovae \citep{alsing2020speculator,kerzendorf2021dalek} and X-ray spectral studies of accretion disk winds in Active Galactic Nuclei (AGN) \citep{matzeu2022new}. Emulators have been built with a wide range of techniques, including: Gaussian processes \citep{bastos2009diagnostics}, artificial neural networks (ANNs) \citep{himes2022accurate}, and kernel ridge regression\citep{vicent2018emulation}. In this paper, we utilize an ANN as an emulator.

ANNs have been shown to obey the universal approximation theorem \citep{baker1998universal}: according to which an arbitrary continuous function of a bounded set can be, within a given accuracy, approximated by an appropriate ANN. As such, they can straightforwardly model even very complex relationships between parameters and model outputs while being fast to compute and extremely memory-efficient in comparison with interpolation methods.

An ANN emulator is trained on a large training data set consisting of pairs of parameters and the corresponding outputs of the model to be emulated. Posterior sampling  of one spectrum may take several million evaluations of the model. Using the same number of model evaluations to generate training data saves time in the long run, as the ANN model, once trained, is orders of magnitude more efficient than the original model. The ANN can also be easily loaded onto a GPU, and model evaluations can be parallelized to achieve even faster evaluation times. 

A reduced model evaluation time allows for faster optimization, sampling and a more comprehensive exploration of parameter space. The latter is especially important to ensure robustness of the statistical inference. Reporting posteriors enables us to report degeneracies and correlations in parameter space. The differentiable nature of a neural network emulator also introduces the possibility of using modern, more efficient methods of optimization and sampling, developed outside of the field \citep{buchner2023nested,girolami2011riemann,cranmer2020frontier}.

This work presents an emulator called \texttt{RTFAST-spectra} (subsequently \texttt{RTFAST}) as a drop-in replacement for the spectral portion of the \texttt{RTDIST} model (expanded upon in section \ref{rtdist}) which is considerably faster than the original model. We outline our methodology in section \ref{RTFAST}, including dimensionality reduction via Principal Component Analysis (PCA), the neural network architecture and a new, custom loss function to efficiently train the network on the PCA representation. In section \ref{Robust}, we present a series of tests of \texttt{RTFAST} to demonstrate its reliability. In section \ref{conclusions}, we summarize our model and discuss its possible use cases and future work.

\section{RTDIST} \label{rtdist}
\texttt{RTDIST} is part of the \textsc{reltrans} package \citep{ingram2019public} and that the new version of \textsc{reltrans} that includes \texttt{RTDIST} will soon be publicly available. A comprehensive introduction to \texttt{RTDIST} and its assumptions can be found in I22. \texttt{RTDIST} incorporates a considerable amount of theoretical development in X-ray reverberation and black hole modelling over the last decade. As such, we provide here a concise overview of \texttt{RTDIST}.

\texttt{RTDIST} assumes a lamppost point source corona emitting a Comptonization spectrum located above a black hole, which has a thermally emitting disk with a fixed aspect ratio of height over radius. The spectrum emitted by the corona is calculated using the thermal Comptonisation model \texttt{nthcomp} \citep{zdziarski1996broad}. The power law index of this spectrum is assumed to vary with time (a feature inherited from other \textsc{reltrans} models), motivated by empirical constraints from observations \citep{kotov2001x}. The spectral pivoting is implemented using a first order Taylor expansion to linearise the model. This emission from the corona travels not only to the observer, but also illuminates the disk. When emission from the corona hits the disk, it is absorbed and reprocessed by material in the disk. This results in re-emission from the disk that is referred to as the reflection spectrum. This reflection spectrum has three key components that are apparent in the observed spectrum: the soft excess below 1keV, the iron line complex around 6.4keV and the Compton hump \citep{ross2005comprehensive,garcia2013x}.

\myRed{This reflection spectrum is also affected by relativistic effects due to the corona and reflecting material being situated near the black hole \citep{fabian1989alignment}. These are modelled via ray-tracing that takes into consideration boosting from directed emission from the corona (rather than simply assuming isotropic emission), as well as blue- and red-shifting effects. This results in relativistic smearing of the entire reflection spectrum, which is most apparent in the narrow line Fe K complex.}

\myRed{Distinct from other models (with the exception of the \texttt{KYN} package; \citealt{Dovciak2004}), the ionisation of the disk is calculated self-consistently from the distance to the source. The ionisation of the disk affects the shape of the resultant reflection spectrum which, combined with the observed flux of the spectrum, allows for the constraining of distance to the source. In conjunction with an observed red-shift, one can theoretically constrain the Hubble constant from the spectra alone with a mass prior from observations in other wavelengths. Including the timing information into the model allows us to measure the distance even without a prior on the mass, since the mass becomes a parameter measured by the model. }

\myRed{Further extending this line of reasoning, considering multiple AGN simultaneously allows for the marginalisation over peculiar velocity of individual sources. This allows for better constraints of distance (and thus ionisation) by sharing this common parameter within fits of data. This enables another} independent method of constraining the Hubble constant with X-ray spectral information. This method also serves as a way to directly verify the geometric assumptions made by \texttt{RTDIST} \myRed{if \texttt{RTDIST} agrees with cosmological measurements of the Hubble Constant}. Theoretically speaking, this method can already be performed with observations made of nearby X-ray bright AGN with XMM-Newton (Mitchell et al in prep). Practically, the complex and long fitting process of an individual AGN makes this difficult to do while retaining the self-consistent calculations required to do this at scale. This particular scientific question directly motivated the development of RTFAST, and RTFAST has several design decisions related to this motivation, such as the modelled energy grid and restricted model parameters.

All parameters in \texttt{RTDIST} are listed in Table \ref{table:pars} as well as their ranges. \texttt{RTDIST} is a general model that can be applied to both AGN and X-ray binaries, however we only train \texttt{RTFAST} for AGN so all parameter ranges are restricted to that use case.

\section{RTFAST} \label{RTFAST}

In this section, we describe the framework for the training of \texttt{RTFAST}. We outline the choices we made when designing \texttt{RTFAST} as well as how we arrived at them. This section can be viewed as a way to approach similar modelling problems to those that we encountered.

\subsection{Data generation}

\texttt{RTDIST} can model a wide range of scenarios, ranging from AGN with low-temperature disks to super-Eddington accreting x-ray binaries. To model every scenario, no matter how unrealistic or unlikely, requires an enormous number of training examples. Thus, we constrain the modelling space to a subset of parameters physically meaningful in modelling AGN. We use sampling techniques, PCA and physical parameter constraints to achieve a manageable parameter space for neural network training. In the following, we will describe how we generate pairs of parameter sets $\theta_i$ and the associated model spectrum generated by RTDIST, $y_i$, $i \in N$. $\theta$ is a vector of parameters described in Table \ref{table:pars}.

\subsubsection{Parameter space} \label{pars_space}

\begin{table*}
\centering
\begin{tabular}{lllll}
Parameter & Description &Spacing & Range & Units                  \\
\hline
h         & lampost corona height &log     &1.5 - 100 & $R_g$                     \\
a         & spin &linear  &0 - 0.998 & *                      \\
Inclination       & inclination &log &  1 - 80   & degrees                \\
$R_{\text{inner}}$    & inner disk radius &log    & 1 - 400 & ISCOs                  \\
$R_{\text{outer}}$    & radius of the outer edge of the disk &log    & 400 - $10^5$ & $R_g$                     \\
z         & redshift &linear & 0 - 0.1 & *                      \\
$\Gamma$     & photon index &linear & 1.4 - 3.4 & *                      \\
$D_{\text{kpc}}$      & angular diameter distance &log   & $10^2$ - $10^6$  & kpc                    \\
$A_{\text{fe}}$       & iron abundance of disk  in solar units &log    & 0.5 - 10 & *                      \\
logNe     & electron density of disk &linear  & 15 - 20 & $N_e$: cm$^{-3}$                      \\
kTe       & observed electron temperature &log     & 5 - 500& keV                    \\
$\frac{1}{\mathcal{B}}$     & boost &log& $10^{-2}$ - 10  & *                      \\
Mass      & mass of BH &log     & $10^4$-$10^{11}$ & $M_{\odot}$                 \\
Scale height      & aspect ratio of conical disk &linear & 0-0.176 & *                      \\
b1        & angular emissivity parameter &linear  & 0-2 & *                      \\
b2        & angular emissivity parameter & linear  & -4 - 4& *                      \\
$A_{\text{norm}}$     & normalisation of the total model &log     & $10^{-4}$ - 4& *                     
\end{tabular}
\caption{List of parameters in \texttt{RTDIST}, the type of parameter space used in training and their units. All units indicated with a * are dimensionless. Note that $A_{\text{norm}}$ is a range calculated from a possible range of coronal fluxes specified in section \ref{phys_prior}.}
\label{table:pars}
\end{table*}

Many of the parameters of \texttt{RTDIST} span multiple orders of magnitude. We use the logarithm of those parameters in training data generation to ensure that all orders of magnitude are equally sampled and represented in the training data. In the case of both very small and very large numbers, this also ensures numerical stability within the neural network. In Table \ref{table:pars}, we list each parameter, the type of parameter space used for that parameter, and its units. In actual usage of \texttt{RTFAST}, users input parameters in the conventional physical units of the original model.

Grid sampling is not very effective in higher dimensions due to the curse of dimensionality \citep{bellman1966dynamic}, and requires large amounts of data to effectively sample the entire parameter space. We used the implementation of Latin Hypercube sampling \citep[LHS;][]{mckay2000comparison} from the \textsc{scipy} python package \citep{2020SciPy-NMeth} to effectively sample the entire parameter space while preventing over/under sampling of areas of parameter space that can occur when uniformly sampling high-dimensional spaces. Latin hypercube sampling achieves this by considering the potential parameter space as components divided into strata of equal marginal probability. Each of these strata are then sampled once and are then randomly matched together. This ensures the entire space is covered effectively and all areas of the overall distribution are explored. This sampling results in all instances of parameter sets $\theta_i$.

\subsubsection{Physically motivated parameter constraints} \label{phys_prior}

\begin{figure}
    \centering
    \includegraphics[width=0.99\linewidth]{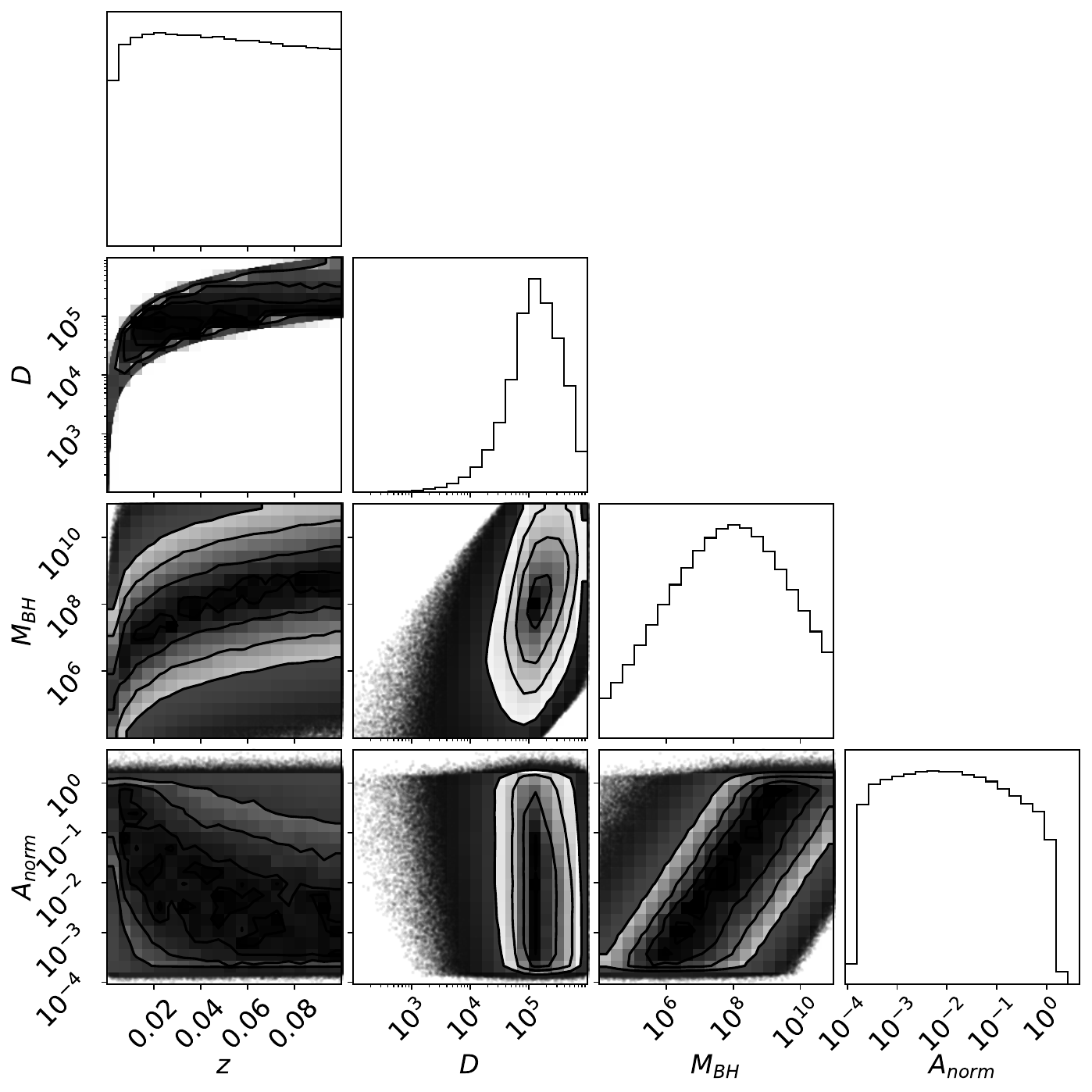}
    \caption{Corner plot of 4 of the training set parameters and their distributions. Possible training data to be generated is constrained by the physical constraints outlined in section \ref{phys_prior}. The full distributions can be found in Fig. \ref{fig:full_phys_prior}.}
    \label{fig:phys_prior}
\end{figure}

Simulating spectra with completely random parameter combinations within the very broad limits set in Table \ref{table:pars} resulted in the simulation of non-physical spectra, such as spectra with emission lines that exceeded the disk continuum emission by 10 orders of magnitude. This introduced additional complexity for the emulator to learn without any scientific gain. We imposed a series of physically motivated parameter constraints that limited our training space to an area that is physically realistic and useful to observers. We imposed 6 physical restrictions, which are the following:

\begin{enumerate}
    \item The redshift and angular diameter distance to the object must be consistent with a Hubble constant between $30$ and $300$km/s/Mpc. Current measurements indicate the Hubble constant to be between $67$ and $73$km/s/Mpc at their most extreme \citep{freedman2021measurements}. We allow for a wider range of possible inferred values to allow for flexibility in inference of the implied ionization of the disk in the final model.
    \item The height of the corona, $h$, must be greater than the horizon radius as otherwise photons emitted from the corona do not escape the black hole.
    \item The coronal flux of the object must be within $10^{-8}$ and $10^{-12}$ erg/s/cm$^{2}$ at earth, the range currently observed from AGN.
    \item The coronal flux of the object must be consistent with an accretion luminosity between $10^{-4}$ and $2.5$ $L_{edd}$ (Eddington luminosity). This aligns with X-ray fluxes of local AGN.
    \item All 3 of the following may not be true simultaneously: 
    \begin{itemize}
        \item the photon index of the corona is greater than 2.75
        \item the disk has an iron abundance of 4 solar units or greater
        \item the electron density of the disk is $10^{19}$ or higher.
    \end{itemize}
    These particular restrictions incorporate knowledge from observations of X-ray binaries where these properties are at their most extreme \citep{ludlam2020nicer,garcia2018reflection,dong2020spin}. Disk densities have been shown to be anti-correlated with black hole mass and higher disk densities have also been shown to decrease the inferred iron abundance of disks of AGN \citep{jiang2019high}, meaning these restrictions serve as a suitable upper limit for AGN.
    \item The spin is forced to be pro-grade since no observational evidences of retrograde spin have been found \citep{Draghis2024}.
\end{enumerate}

These restrictions are applied to all parameter sets $\theta_i$ generated in section \ref{pars_space}. This functionally necessitated the removal of instances of $\theta_i$ that were invalid, resampling the parameter space, and checking if these new parameter sets were valid. This was repeated until the desired number of parameter sets $\theta_i$ was reached.

These physical constraints are also visualised in Fig.~\ref{fig:phys_prior}. The full distributions can be found in Appendix \ref{full_dist}. By imposing these physical restrictions, we reduced the possible parameter space to $1\%$ of its original size (Table \ref{table:pars}) and eliminated non-physical results. Fig. \ref{fig:phys_prior} shows clear correlations between parameters: for instance, $A_{\text{norm}}$ and mass show a dependence such that increasing $A_{\text{norm}}$ only allows for increasingly smaller range of masses: this is due to $A_{\text{norm}}$ being highly correlated with accretion luminosity and coronal flux. Lower values of $A_{\text{norm}}$ mean that AGN with smaller masses do not produce enough flux to be meaningfully observed at earth at the distances specified by our physical constraints. This effect is also seen in the correlations between $A_{\text{norm}}$ and redshift (and subsequently distance).

We note that, due to these restrictions, moving outside of these imposed physical bounds is likely to give incorrect results. Any usage of \texttt{RTFAST} should impose these physical constraints. We demonstrate this restriction in section \ref{outofbounds}. If observations implied scenarios outside of these parameter constraints, it would be relatively trivial to loosen these restrictions by re-training the neural network on new data outside the original constraints.

\subsubsection{Pre-processing of spectra} \label{pre-processing}

Neural networks training performs best when the values the network is supposed to reproduce are well-behaved and narrowly distributed. Therefore, we applied several reversible pre-processing steps. The neural network learns a relationship between the parameter and these pre-processed quantities, which can then be easily and efficiently expanded into a physically meaningful spectrum.

\begin{enumerate}
    \item All simulated fluxes below $10^{-11}$ photons/cm$^2$/s/keV in the spectrum were fixed to $10^{-11}$ photons/cm$^2$/s/keV. This prevents the emulation of fluxes that we realistically will not observe. This threshold corresponds to an average of one photon per energy bin detected during a 1 Megasecond observation with the Athena X-IFU telescope \citep{Kammoun2022}. This is realistically extremely conservative and no current or planned telescope at the time of writing will detect photons below this limit.
    \item Neural networks are known to struggle generating outputs that span multiple orders of magnitude. To mitigate this issue, we took the logarithm of the flux $y_i$ in each energy bin $E_i$ to obtain a new log-scaled quantity, $x_i = 
    \log(y_i)$. 
    \item Each $x_i$ is then scaled such that the mean of $x_i$ for each bin $E_i$, taken over all $N$ training examples, is 0 and the standard deviation is 1. This scaling enhances features such as changes in the iron absorption and reflection lines that deviate significantly from the average spectrum, both of which are particularly important in our science case.
    \item We performed a PCA decomposition with 200 PCA components on the spectra. We go into more detail about this technique in section \ref{PCA}.
    \item The 200 PCA components are scaled to a mean of 0 and standard deviation of 1 over the dataset for identical reasons as in step 3, with the additional motivation that neural networks behave more stably when the target output is constrained around 0 with magnitudes of approximately 1.
\end{enumerate}

This pre-processing was key in achieving and maintaining high precision over the entire spectrum by reducing the complexity of the model space and reducing the size of the neural network required to model the spectral behaviour.

\subsubsection{PCA decomposition} \label{PCA}

\begin{figure*}
    \centering
    \includegraphics[width=0.99\linewidth]{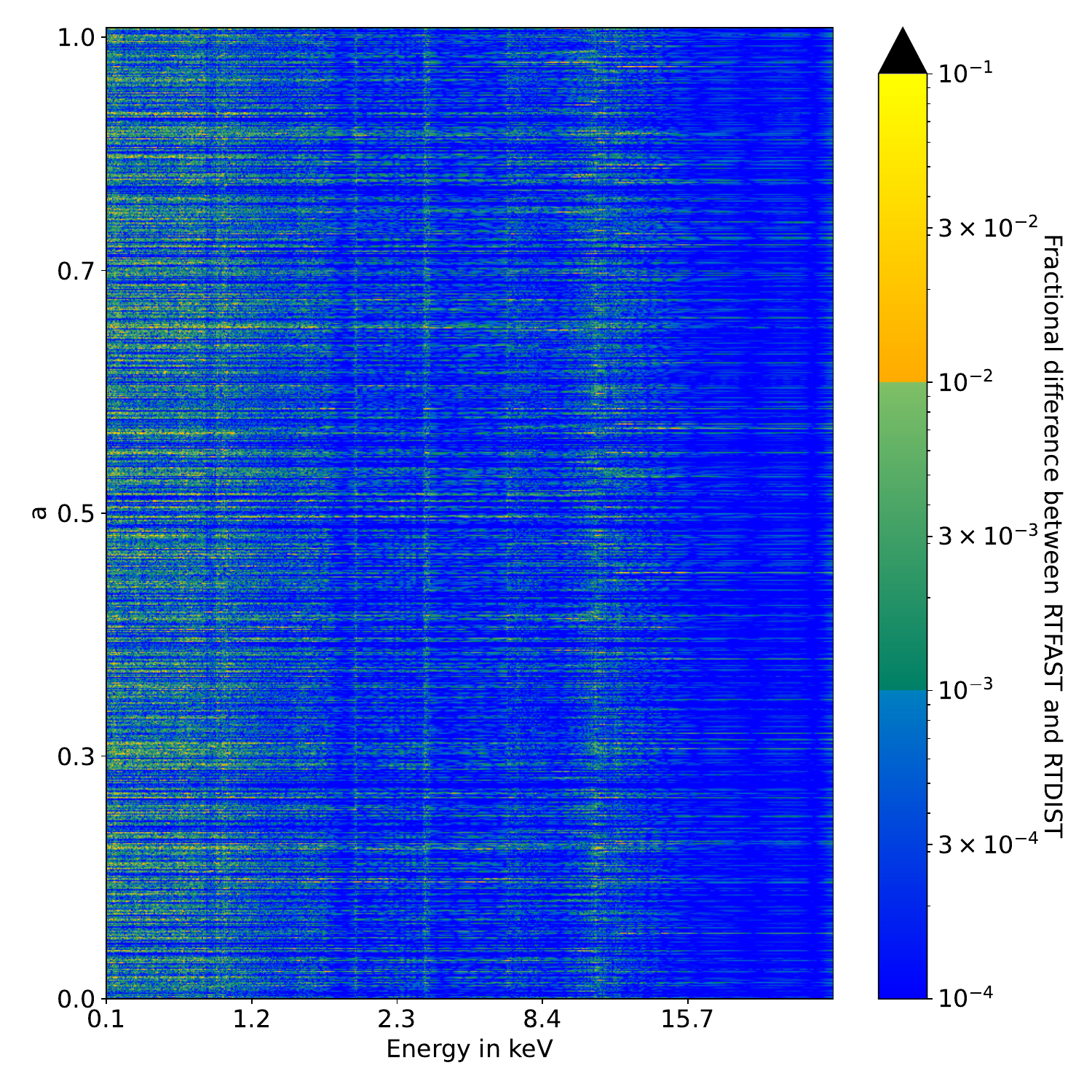}
    \caption{A heat-map of the percentage residual difference between the PCA reconstruction and \texttt{RTDIST} sorted by the model parameter spin (a). Yellow to orange indicates $1$-$10\%$ error, green indicates $0.1$-$1\%$ error and blue indicates $0.01\%$ error and below. Black indicates an error above $10\%$. White indicates that both the \texttt{RTFAST} and \texttt{RTDIST} output are below the detectable limit outlined in section \ref{pre-processing}.}
    \label{fig:PCA}
\end{figure*}

The large number of energy bins present in X-ray spectra requires large neural networks by design. In turn, these networks struggle to capture small sharp features such as the iron line, as most spectral bins are dominated by the continuum emission, which thus dominated training performance. To reduce the number of outputs the neural network has to predict, we performed PCA decomposition to capture the relevant structure in the spectra in a lower-dimensional space\footnote{\citet{alsing2020speculator} is a historical example of using PCA decomposition to enable better emulation of spectra.}. This reduced the spectrum from 2017 energy bins (the XMM-Newton EPIC-PN energy grid above 0.1~keV) to $K$ principal components. We found that by decomposing the spectra, correlations between energy bins were effectively preserved. This also led to considerably smaller neural networks as it reduced the problem to learning a relationship between 17 input parameters and K output features, which thus significantly reduced training time and complexity. 

In order to determine $K$, we need to balance the trade-off between the reduction of the dimensionality of the output space to be predicted by the neural network with the loss of accuracy implied by the use of a finite set of PCA components. To do this, we iteratively increased the number of PCA components $K$, and computed the average error in the reconstructed spectra across a set of test spectra that were not used in computing the PCA components. We stopped adding components when this average error fell below a defined threshold. We chose to set this threshold at $0.1\%$ over the entire parameter space, and defined the error as 

\begin{equation} \label{eq:resid}
    \chi_i = (x_{i, \mathrm{NN}}-x_{i, \mathrm{RTDIST}})/x_{i, \mathrm{RTDIST}}
\end{equation}
\noindent where $x_{i, \mathrm{NN}}$ is the spectrum in bin $i$ reconstructed from the PCA components, transformed as described in steps (i) - (iii) in Section \ref{pre-processing}, $x_{i, \mathrm{RTDIST}}$ and denotes the corresponding transformed spectrum generated with RTDIST. This equation describes the fraction error $\chi_i$, which we will also use in later on to assess the reconstruction performance of the neural network combined with PCA.

At a threshold of 0.1\% reconstruction error, we effectively require that the PCA reconstruction must be an order of magnitude better than the performance we require from the neural network. 

We found that $K=200$ PCA components generally fulfilled our requirement, as shown in Figure \ref{fig:PCA}.

\begin{figure}
    \centering \includegraphics[width=0.99\linewidth]{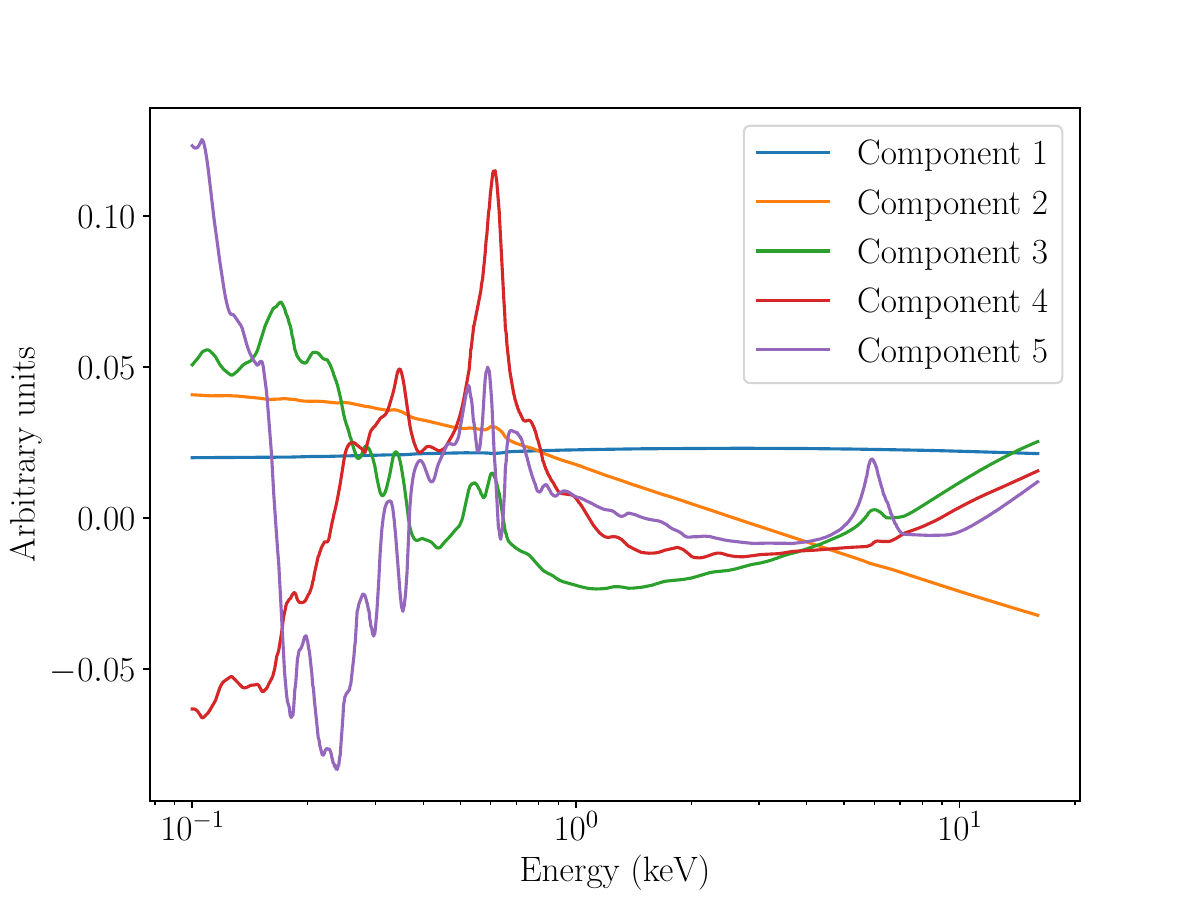}
    \caption{The first 5 PCA components plotted as a function of energy. Each component is in arbitrary units and is scaled appropriately for their actual contribution in the final spectrum.}
    \label{fig:pca_comps}
\end{figure}

\begin{figure}
    \centering \includegraphics[width=0.99\linewidth]{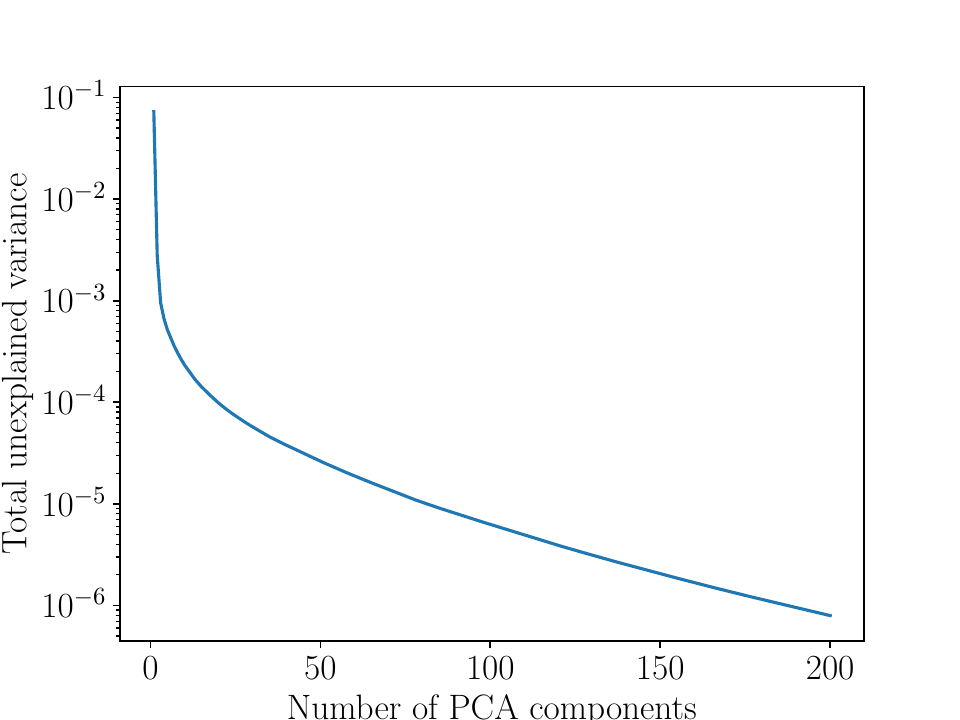}
    \caption{Cumulative summing of the explained variance by PCA component subtracted from 1.}
    \label{fig:explained_variance}
\end{figure}

Fig. \ref{fig:pca_comps} shows the first 5 PCA components plotted as a function of energy. The original spectrum is recreated from a linear combination of these components, weighted by the PCA projection of the original data. It is these PCA projections that the neural network learns to emulate. Because PCA is a reversible operation, reconstructing the original spectra from the projections is trivial.
The first component in Figure \ref{fig:pca_comps} is largely flat and thus encapsulates the effect of a normalising constant\footnote{Ordinarily, we would divide out a normalising constant from the spectrum and add it as a separate output to be modelled by the neural network. However, for RTDIST, the normalisation is not independent from the shape of the spectrum, thus this route is not available to us.}. The second PCA component is dominated by a power-law like feature. 

\myRed{Notably, in Fig \ref{fig:pca_comps}, components 3 and onwards exhibit much more interesting features than the simple normalisation and power law component that the first 2 components can largely explain. Later components can be directly related to the soft excess below 1keV and the iron line complex at 6.4keV. These components show how the PCA decomposition directly captures physical features of interest. The remaining components that are not shown in this plot almost exclusively relate to better describing the soft excess and iron line.}

\myRed{Fig. \ref{fig:explained_variance} shows the cumulative explained variance ratio subtracted from 1. The explained variance ratio is a statistical measure of how much of the variation of a dataset is captured by each principal component. If the explained variance ratio sums to 1, all variance in the dataset is explained, and any input from the dataset can be perfectly reconstructed. The first two components capture the majority of the explained variance, accounting for $99.7\%$ of the variance in the model. The following components capture increasingly smaller amounts of the overall explained variance despite being very important to recreating features in the spectrum such as emission lines.}

\subsubsection{Dataset split} \label{data_split}

We generated a total of $1.1\times10^7$ spectra using the \textsc{Sherpa} python package \citep{freeman2001sherpa} to interface with the original \texttt{RTDIST} model on the energy grid of the XMM-Newton PN CCDs from 0.1~keV to 20~keV. This dataset amounts to a total of approximately 700GB. We split the data into a training set and validation set with a $90\%$\textbackslash $10\%$ split. We found a reliable increase in model accuracy with increasing of dataset size, but did not increase the training dataset size further as we reached the $\mathcal{O}(1\%)$ accuracy requirement while also approaching storage limits on the computer we trained on. We also separately generated a test set of 1000 spectra for performance review using the same methodology as the training and validation set.

\subsection{Neural network} \label{training_full}
A neural network consists of a large ensemble of neurons, arranged into several layers and nodes. The output of the $m^\mathrm{th}$ node in the $n^\mathrm{th}$ layer is $a_m^n = \sigma( \mathbf{a}^{n-1} \odot \mathbf{W}_m + \mathbf{b}_m $), where $\mathbf{a}^{n-1}$ is the output of the $(n-1)$th layer, $\mathbf{W}_m$ and $\mathbf{b}_m$ are trainable arrays of hyper-parameters, and $\sigma()$ is the activation function.

Training of the neural network involves optimising the parameters to minimise a loss function. This optimisation is achieved using stochastic gradient descent, which allows for the stochastic approximation of gradient descent optimisation.

We explored a variety of strategies to train \texttt{RTFAST} including: loss functions, different optimizers, learning rate schedulers and architectures. We ultimately achieved best performance with a somewhat conventional approach with the exception of a custom loss function. We report how we reached this approach and our best performing strategy in the following section.

\subsubsection{Architecture} \label{architecture}

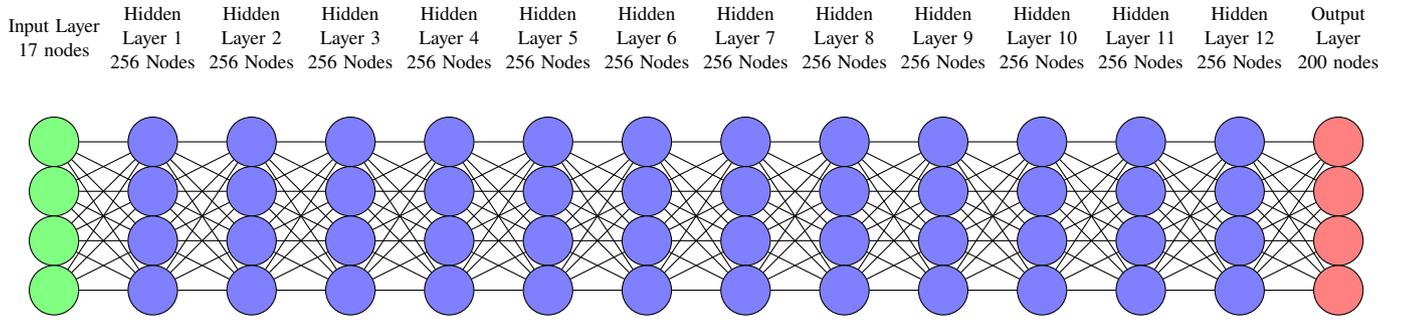
\begin{figure*}
    \begin{tikzpicture}[scale=0.65, transform shape]

    \tikzstyle{neuron}=[circle, draw, minimum size=1cm]
    \tikzstyle{input neuron}=[neuron, fill=green!50]
    \tikzstyle{output neuron}=[neuron, fill=red!50]
    \tikzstyle{hidden neuron}=[neuron, fill=blue!50]
    \tikzstyle{annot} = [text width=5em, text centered, scale = 1.4]
    
    \foreach \name / \y in {1,...,4}
        \node[input neuron] (I-\name) at (0,-\y) {};
    
    \foreach \layer in {1,...,12}
        \foreach \name / \y in {1,...,4}
            \path[yshift=0cm]
                node[hidden neuron] (H-\layer-\name) at (\layer*2,-\y) {};
    
    \foreach \name / \y in {1,...,4}
        \path[yshift=0cm]
            node[output neuron] (O-\name) at (26,-\y) {};
    
    \foreach \source in {1,...,4}
        \foreach \dest in {1,...,4}
            \path (I-\source) edge (H-1-\dest);
    
    \foreach \layer in {1,...,11}
        \foreach \source in {1,...,4}
            \foreach \dest in {1,...,4}
                \path (H-\layer-\source) edge (H-\the\numexpr\layer+1\relax-\dest);
    
    \foreach \source in {1,...,4}
        \foreach \dest in {1,...,4}
            \path (H-12-\source) edge (O-\dest);
    
    \node[annot, above of=H-1-1, node distance=1.5cm] (hl1) {Hidden Layer 1\\256 Nodes};
    \node[annot, above of=H-2-1, node distance=1.5cm] (hl2) {Hidden Layer 2\\256 Nodes};
    \node[annot, above of=H-3-1, node distance=1.5cm] (hl3) {Hidden Layer 3\\256 Nodes};
    \node[annot, above of=H-4-1, node distance=1.5cm] (hl4) {Hidden Layer 4\\256 Nodes};
    \node[annot, above of=H-5-1, node distance=1.5cm] (hl5) {Hidden Layer 5\\256 Nodes};
    \node[annot, above of=H-6-1, node distance=1.5cm] (hl6) {Hidden Layer 6\\256 Nodes};
    \node[annot, above of=H-7-1, node distance=1.5cm] (hl7) {Hidden Layer 7\\256 Nodes};
    \node[annot, above of=H-8-1, node distance=1.5cm] (hl8) {Hidden Layer 8\\256 Nodes};
    \node[annot, above of=H-9-1, node distance=1.5cm] (hl8) {Hidden Layer 9\\256 Nodes};
    \node[annot, above of=H-10-1, node distance=1.5cm] (hl8) {Hidden Layer 10\\256 Nodes};
    \node[annot, above of=H-11-1, node distance=1.5cm] (hl8) {Hidden Layer 11\\256 Nodes};
    \node[annot, above of=H-12-1, node distance=1.5cm] (hl8) {Hidden Layer 12\\256 Nodes};
    
    \node[annot, above of=I-1, node distance=1.5cm] (il) {Input Layer\\17 nodes};
    \node[annot, above of=O-1, node distance=1.5cm] (ol) {Output Layer\\200 nodes};
    
    \end{tikzpicture}
    \caption{Network architecture of \texttt{RTFAST}. \texttt{RTFAST} is composed of 12 hidden layers, each with 256 nodes, an input layer of 20 nodes and an output layer of 40 nodes. Between each layer, with the exception of the output layer, is a GELU activation function.}
    \label{fig:architecture}
\end{figure*}

The neural network in this paper is built using the \textsc{Pytorch} package \citep{paszke2019advances}. We utilised a feed-forward neural network. The architecture was deliberately kept methodologically very simple as previous work \citep{rivera2015emulator,kwan2015cosmic} has found high efficacy with feed-forward neural networks compared to those with more sophisticated architectures. Smaller networks also reduce overhead in computation and complexity in training. 

After preliminary tests on a network that was hand-tuned, we initially performed grid sweeps of different architectural structures during training: changing the number of layers (4, 6, 8, 10, and 12), nodes (256, 512, and 1024) and type of activation function (Rectified error Linear Unit (ReLU), and Gaussian Error Linear Unit (GELU)) used in the network to find the best performing model while keeping learning hyper-parameters fixed. ReLU activation functions can be defined as the non-negative part of its argument, also known as the ramp function. This is defined as:

\begin{equation}
    \text{ReLU}(x) = x^+ = \text{max}(0,x) = \frac{x+|x|}{2}
\end{equation}

GELU activation functions can be considered a smoother form of the ReLU that are widespread in neural networks in the literature today. The GELU activation function is $x\Phi(x)$, where $\Phi(x)$ is the standard Gaussian cumulative distribution function. GELU weights inputs by their percentile rather than gating the inputs by their sign like in ReLU. The $\text{GELU}(x)$ is defined as as:

\begin{equation}
    \text{GELU}(x) = x\Phi(x) = x\cdot \frac{1}{2}[1 + \text{erf}(x/\sqrt{2})]
\end{equation}

We found that a network of 8 fully connected linear layers with 256 nodes in each layer with GELU activation functions \citep{hendrycks2016gaussian} between each hidden layer to perform the best for our use case. We show a visual representation of the best performing network in Fig \ref{fig:architecture}.

\subsubsection{Hyper-parameter grid sweeps} \label{grid_sweeps}

We then performed grid sweeps of hyper-parameters, including: batch size (256, 512, 1024), learning rate ($10^{-4}$, $5\times10^{-3}$, $10^{-3}$), optimizers (SGD \citep{sutskever2013importance}, Adam \citep{kingma2014adam}, and AdamW \citep{loshchilov2017decoupled}) and learning rate schedulers (no scheduling, Cosine annealing \citep{loshchilov2016sgdr}, and ExponentialLR). Grid sweeps in machine learning are the systematic training of a neural network with different neural networks and sets of pre-specified hyper-parameters (where hyper-parameter is a parameter that defines the details of the learning process of a neural network) to then compare final performance and select the best performing combination. Hyper-parameters include (but not exclusively): learning rates, the type of optimizer used during training, and batch size. This approach often results in faster development than trial and error development of a neural network at the cost of computational power. 

We found that a learning rate of $10^{-4}$, a batch size of 1024, an Adam optimizer with otherwise default hyper-parameters and no learning rate scheduler yielded the best results. In general, we found larger batch sizes and lower learning rates to perform better. The addition of scheduling was negative for the Adam and AdamW optimizers.

\subsection{Training} \label{training}
During training, we used a custom loss function that took into account the relative importance of each PCA component. We used a weighting factor to push the network to prioritise learning components that had the largest effect on the final reproduced spectrum. This also made the neural network training process more stable when compared to the non-weighted scenario. 

We achieved this by weighting each component by a factor $w_i$ which was calculated as the following:
\begin{equation}
    w_i = \log_{10}(\rm{var}(PC_i)/\rm{var}(PC_n)) + 1
\end{equation}
where $\rm{var}(PC_i)$ and $\rm{var}(PC_n)$ denote the explained variance of the principal component to be weighted and the least informative principal component respectively. The additional constant of 1 ensures that no component is totally ignored during training. The final loss function takes the form:
\begin{equation}
    L = \sum_{i=0}^{n} w_i \cdot (PC_i - \hat{PC}_i)^2
\end{equation}

where $PC_i$ is the scaled principal component, $\hat{PC}_i$ is the neural network output. This can be considered a weighted mean square error loss.

\begin{figure}
    \centering
    \includegraphics[width=0.99\linewidth]{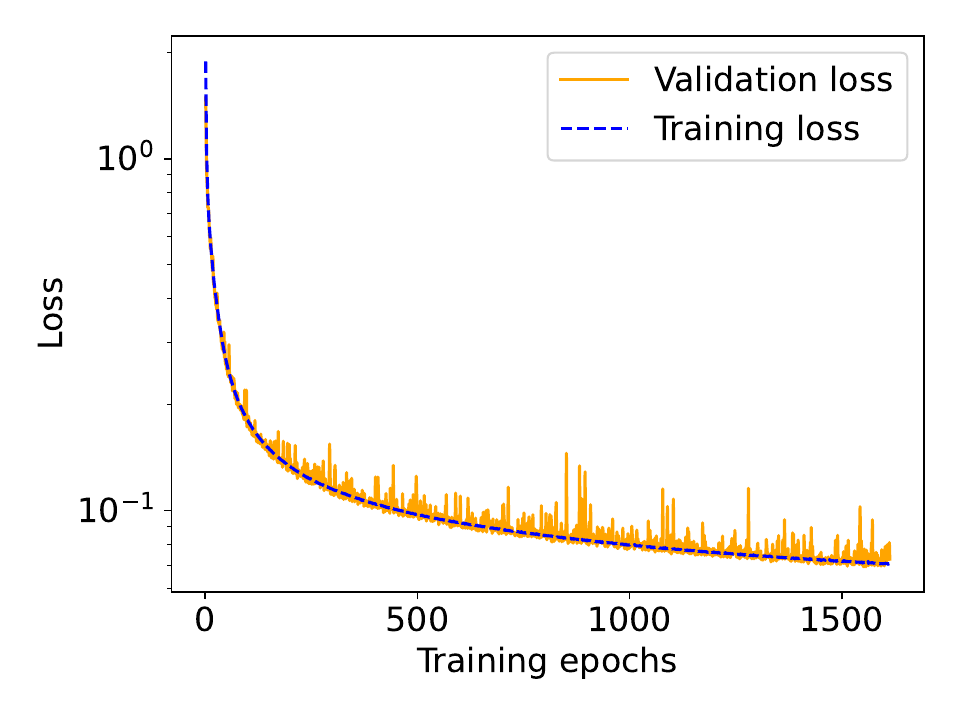}
    \caption{Training and validation loss as a function of training epoch. Due to the large training data size, initial validation loss is lower than training loss.}
    \label{fig:loss}
\end{figure}

We trained all neural networks described in sections \ref{architecture} for a maximum of 2000 epochs (where an epoch is one complete pass of the training set through the algorithm) with an early stopping criterion of requiring at least $1\%$ improvement over 100 epochs. In fig. \ref{fig:loss}, we show the training and validation loss as a function of training epoch for the best performing network found by performing grid sweeps as described in sections \ref{architecture}. While both training and validation loss continued to improve over the last 100 epochs, we found the two were improving increasingly slowly, indicating that the network had learnt most of the information in the dataset.

\subsection{Ensemble}
\begin{figure}
    \centering
    \includegraphics[width=0.99\linewidth]{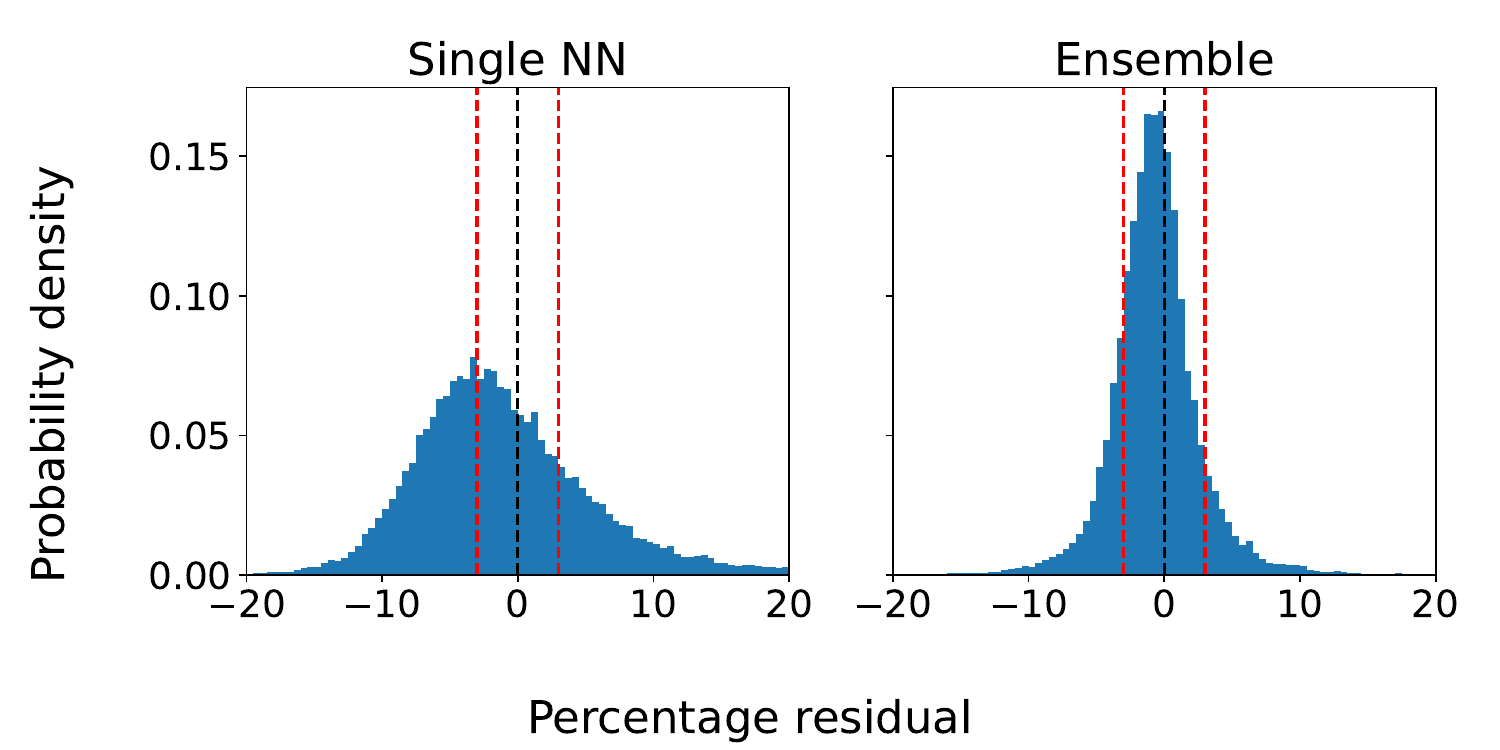}
    \caption{Comparison of residual probability density functions on the test set between using a single neural network and the ensemble. The dashed red vertical lines indicate $1\%$ error and the dashed blue vertical line indicated $0\%$ error. The ensemble centers the emulator outputs directly around the truth, largely eliminating systematic over- or under-estimation.}
    \label{fig:calib}
\end{figure}

During training of the network, stochastic gradient descent optimization moves the network towards the global minimum of the loss landscape. While theoretically possible to reach the global minimum, this often takes a prohibitively long time and training is stopped before this global minimum is reached. A computationally less expensive strategy to improve performance of a neural network prediction is through the use of an "ensemble" of neural networks \citep{ganaie2022ensemble}. Each of the neural networks in an ensemble is trained separately either on the full dataset or on a subset of it. Each of these neural networks will be trained to try and predict the correct solution but will give a slightly different solution to a given set of parameters. Averaging the outputs of each of the neural networks together has been shown to yield improvements in accuracy in regression problems like emulation \citep{mendes2012ensemble,lakshminarayanan2017simple}. 

We trained 7 instances of the best performing neural network architecture from section \ref{architecture}, following the same training pattern as in section \ref{training}. The final validation losses of each of these networks were comparable to one another, indicating similar levels of convergence as for the initial network (Fig. \ref{fig:loss}). 

We compared the improvement yielded from considering the ensemble of neural networks to a single instance of the neural network.
To evaluate performance, we use the same fractional residual differences between the true spectrum and the prediction from \texttt{RTFAST}, parametrized in Equation \ref{eq:resid}, as for the PCA reconstruction. In Section \ref{PCA}, we only compared how well PCA reconstructed the original spectra in order to understand the number of components required. Here, we compare the original spectra generated by \texttt{RTDIST} to emulated spectra, generated by letting the ANN predict PCA components from the same input parameters, which are then transformed as described in Section \ref{pre-processing} into spectra comparable to \texttt{RTDIST}'s output.
For the test set, we obtain a distribution of residual differences per energy bin. Ideally, this distribution would be a delta function centred at 0 indicating that \texttt{RTFAST} perfectly reproduces the \texttt{RTDIST} output. In reality, this is not the case. In Fig \ref{fig:calib}, the left hand panel shows the distribution of residuals for a single neural network while the right hand panel shows the distribution of residuals for the ensemble. What is apparent is that the single neural network overestimates the true output on average by about $3\%$ and exhibits a skewed distribution while the distribution of the ensemble instead is centered much closer to 0, implying that the ensemble on average estimates the correct output to better precision. The distribution of errors of the ensemble is tighter, with approximately $56\%$ more of the residuals sitting within the bounds of $-1$ to $1\%$.

While evaluating 7 neural networks does increase computation time, use of vectorisation means that the total evaluation time of the ensemble is only approximately $2.7$ times longer than evaluating a single neural network and the mitigation of error and systematic bias is worth the increase in computation time. We adopt this ensemble averaging method for the final version of RTFAST.

\subsection{Computation speed} \label{speed}

A key goal of \texttt{RTFAST} was to speed up model evaluation. \texttt{RTFAST} takes $1.62\times10^{-3}$ seconds for a single set of parameters on a Intel Xeon Gold 6342 2.80GHz processor. This is a 200 times speed up when directly compared to the $0.32$ seconds it takes to call \texttt{RTDIST} on the same processor. This is the most conservative estimate of the speed up as the neural network also allows for vector computation, enabling parallel computation of models at much higher speed than in serial. For instance, the computation of 1000 different models on \texttt{RTFAST} takes $3.56\times10^{-2}$ seconds, while it took \texttt{RTDIST} 314 seconds called sequentially. This represents a speed up of 8000 times faster than \texttt{RTDIST} and $39$ times faster than calling \texttt{RTFAST} sequentially. 

The time it takes to perform a single evaluation of \texttt{RTFAST} in even the most conservative conditions is similar or smaller to the amount of time it takes to perform a single response convolution of XMM-Newton EPIC responses with the model output. This means that parameter inference is no longer computationally limited by the physical model but instead by the convolution of the instrument response and calculation of the likelihood.

\section{Robustness} \label{Robust}

In this section, we perform a suite of tests to assess the robustness of the emulator. We consider the emulator to be robust if it reliably reproduces the true output of \texttt{RTDIST} to within $\mathcal{O}(3\%)$ and without any clear systematic biases. As we envision using \texttt{RTFAST} with XMM-Newton data, we use the systematic error of the XMM-Newton instrumentation as a threshold of how far off the truth \texttt{RTFAST} can be. The most recent cross-calibrations estimate the MOS and PN cameras to have an error in relative effective area to be within $\pm3\%$ and $\pm2\%$ respectively \citep{guainazzi2013xmm}. We subsequently adopt a criterion of $3\%$ error. We explored multiple ways to show that \texttt{RTFAST} fulfills these criteria as well as performing fits of simulated observations from \texttt{RTDIST}. 

\subsection{Direct comparison} \label{compare}

\begin{figure}
    \centering
    \includegraphics[width=0.99\linewidth]{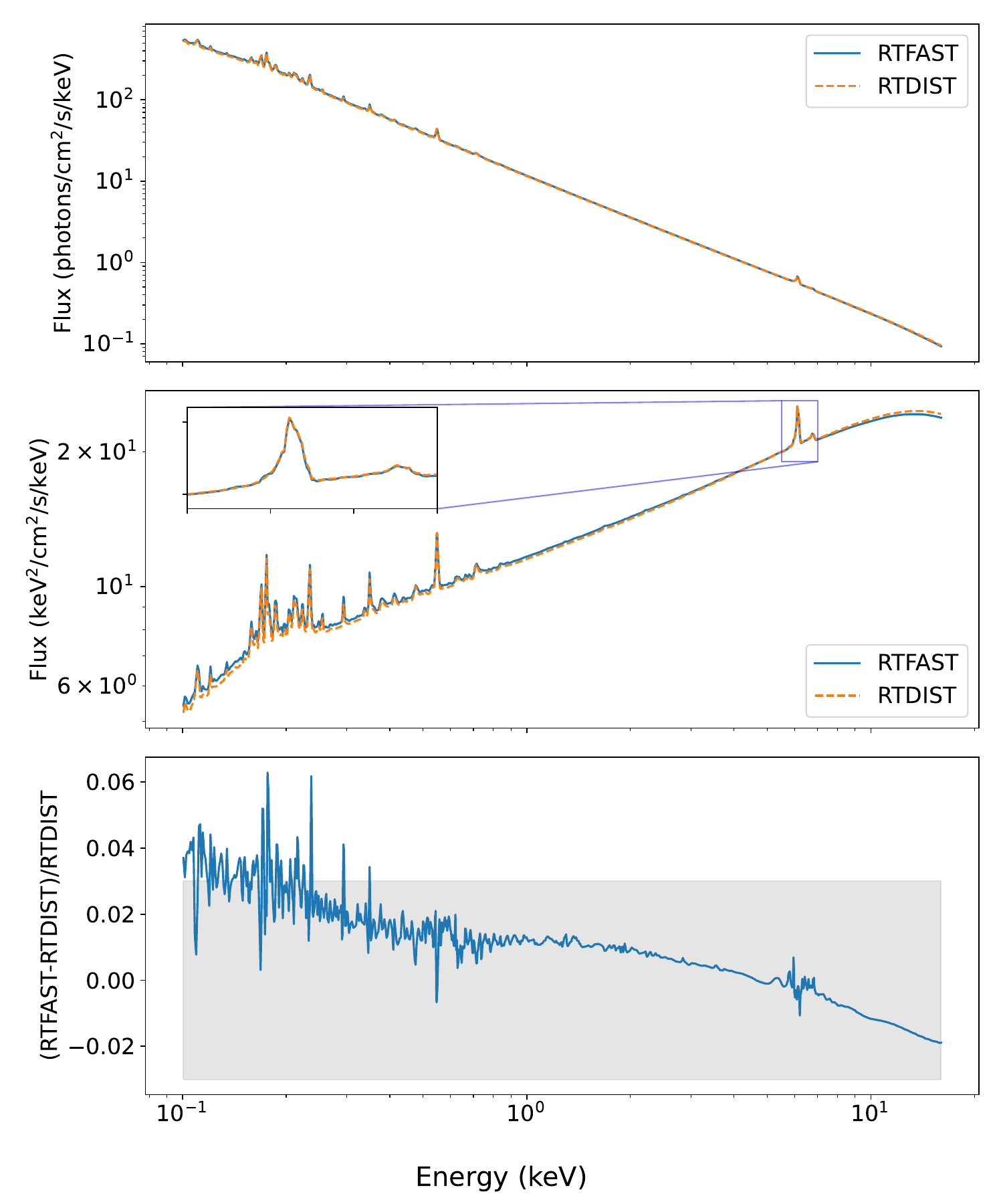}
    \caption{A direct comparison of \texttt{RTFAST} and \texttt{RTDIST}. The top panel shows the photon flux spectrum from \texttt{RTFAST} and \texttt{RTDIST} as a function of energy, with both lines on top of one another. The middle panel shows the same in ${\rm keV}^2/{\rm cm}/{\rm s}/{\rm keV}$ units. The inset panel shows a zoom-in of the relativistically smeared iron line between 5.5 and 7keV. The bottom panel shows the fractional residual difference between \texttt{RTFAST} and \texttt{RTDIST} as a function of energy. The $3\%$ error region is shaded.}
    \label{fig:compare}
\end{figure}

In Fig. \ref{fig:compare}, we directly compare \texttt{RTFAST} to \texttt{RTDIST} for one example from the test set. We observe that there are differences between the models as a function of energy but note that all of these differences are within the desired accuracy above 1keV. At the soft excess at low energies, the difference between \texttt{RTFAST} and \texttt{RTDIST} becomes larger and more jagged. While this is somewhat of a theme throughout the overall parameter space for the emulator, the soft excess is largely obscured by hydrogen column absorption and detector effects in actual observations. These discrepancies at higher resolutions become largely washed out under realistic observation conditions. We discuss this further in a fit to simulated data in section \ref{fit}.

\subsection{Expected model behaviour} \label{behaviour}

\begin{figure}
    \centering
    \includegraphics[width=0.99\linewidth]{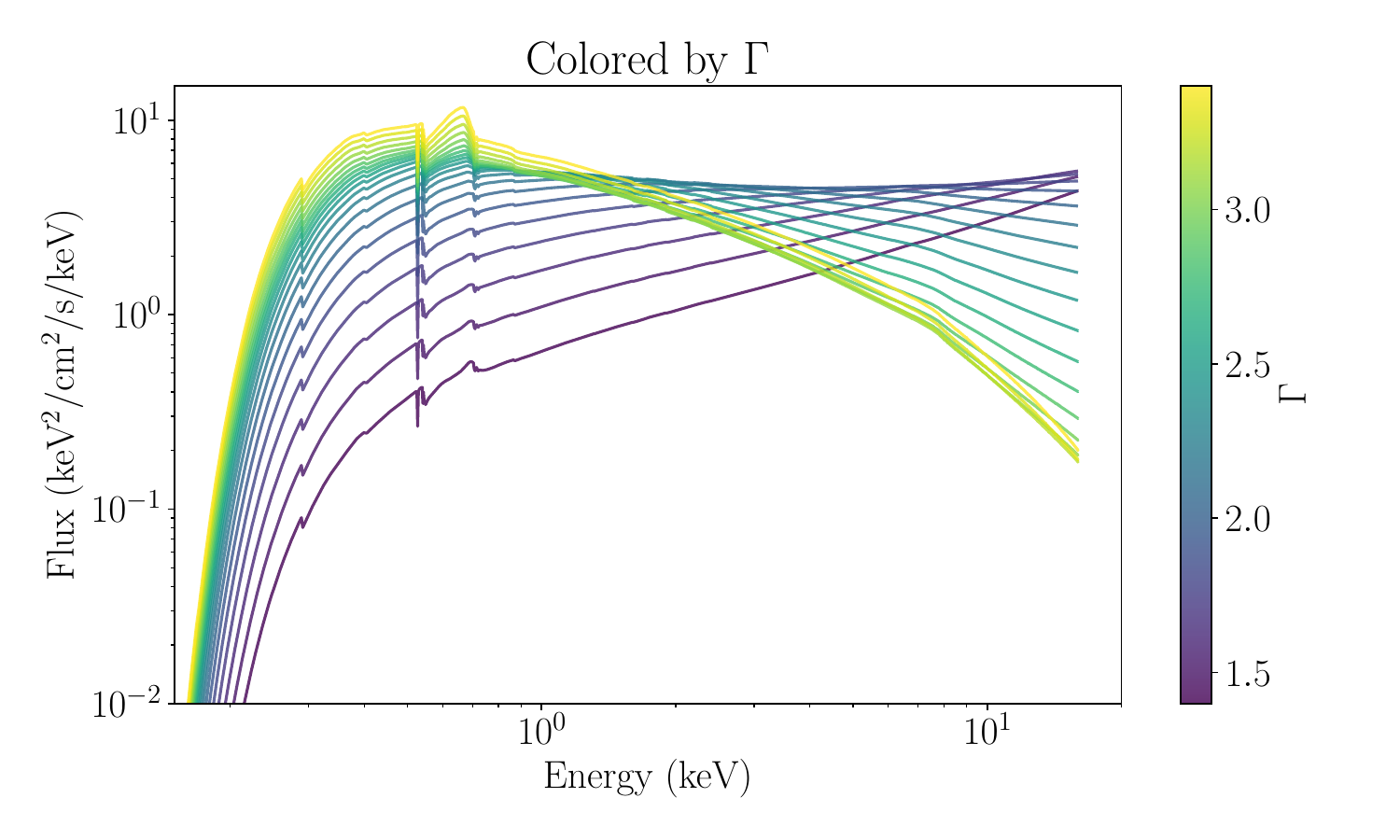}
    \caption{Model spectra generated by \texttt{RTFAST} are plotted on top of one another, coloured by increasing brightness with increasing $\Gamma$ with all other parameters fixed to that of the parameters in Table \ref{table:fits}.}
    \label{fig:Gamma}
\end{figure}

In addition to a direct comparison from a randomly chosen parameter set, we also tested that we observe the expected changes to the spectrum from varying just one parameter. We vary $\Gamma$ due to its easily observable change in the continuum spectrum as well as absorption and reflection lines becoming more apparent as $\Gamma$ is increased. In fig. \ref{fig:Gamma}, we plot the resulting spectra from \texttt{RTFAST} (multiplied by \texttt{TBABS} to model the effects of a line of sight absorption column, using the abundances of \citet{wilms2000absorption}) as a function of $\Gamma$. We observe the expected behaviour of steepening of the spectrum as $\Gamma$ is increased. We also observe the appearance of the Fe K complex within the 6-8 keV range with the increasing steepness of the slope---a key component of reflection modelling within the model and a complex feature to model. Due to the relativistic smearing present in the original model, these are not sharp lines.

Accurately reproducing key emission lines such as the Fe K complex is of key importance to a robust data analysis with the emulator. These lines contain a wealth of physically relevant information that enable tight constraints on model parameters when applied to data. It is not an exaggeration to say that the failure to model these key areas of the spectrum would result in a practical inability to use \texttt{RTFAST} in modelling real observations.

\subsection{Inaccuracy as a function of parameter space and energy} \label{heatmap}

\begin{figure*}
    \centering
    \includegraphics[width=0.99\linewidth]{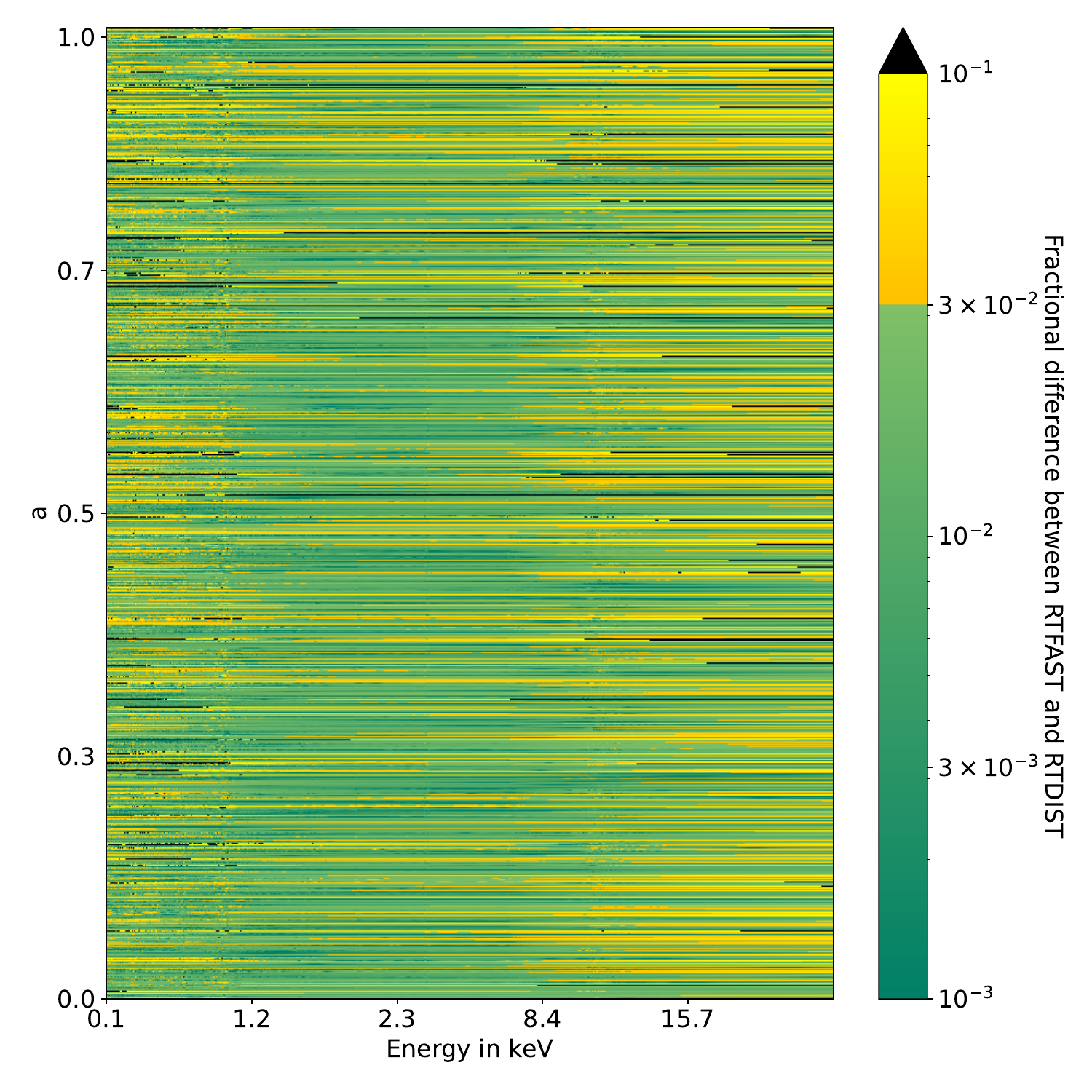}
    \caption{A heat-map of the fractional residual difference between \texttt{RTFAST} and \texttt{RTDIST} (see eq. \ref{eq:resid}). Yellow to orange indicates $1$-$10\%$ error, green indicates $0.01$-$3\%$ error and below. Black indicates an error above $\%$. White indicates that the model output is below the detectable limit outlined in section \ref{pre-processing}.An error of $3\%$ or less is comparable to instrumental error of current X-ray telescopes with the exception of XRISM. At the time of writing, \texttt{RTDIST} is not capable of XRISM resolution, so we did not attempt to achieve such resolution in \texttt{RTFAST}.}
    \label{fig:heatmap}
\end{figure*}

In fig. \ref{fig:heatmap}, we show the fractional difference between \texttt{RTFAST} and \texttt{RTDIST} for the entire test set as a function of black hole spin ($a$). We find that \texttt{RTFAST} generally performs within our requirements of $3\%$ or lower error in over $71\%$ of individual energy bins and extends to $15\%$ in the worst case scenarios. We do not find a consistent trend in increasing error as a function of any parameter, indicating that such errors are stochastically spread around the true model (as indicated previously in fig. \ref{fig:calib} and fig. \ref{fig:heatmap}). However, this means that small individual biases towards over- or under-estimating the spectrum occur in small pockets of parameter space. This impacts the final posterior in the worst scenarios. We discuss this further with an example fit to a simulated spectrum in section \ref{fit}.

\begin{figure}
    \centering
    \includegraphics[width=0.99\linewidth]{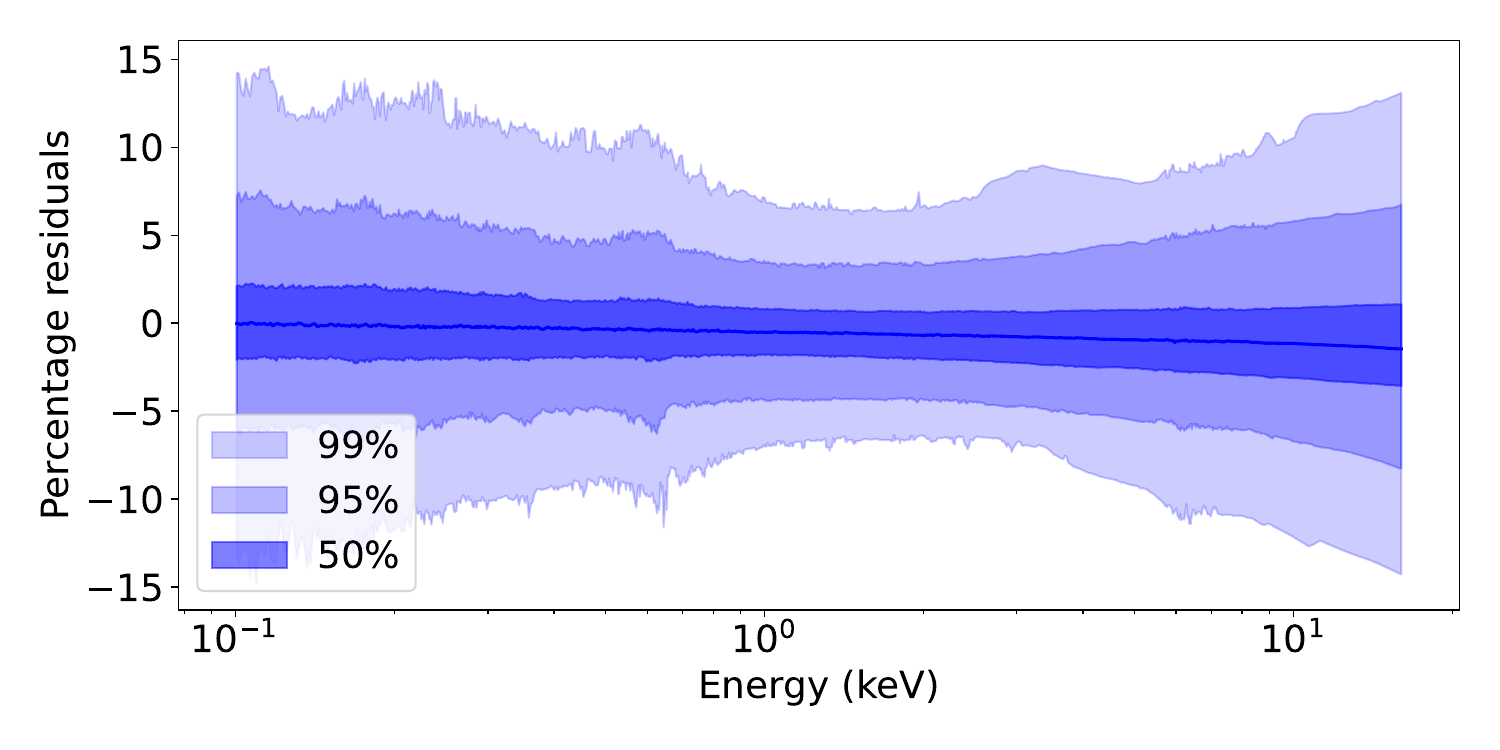}
    \caption{Distribution of the percentage residual difference between \texttt{RTFAST} and \texttt{RTDIST} as a function of energy. The shade of region indicates the $25-75\%$, $5-95\%$, and $1-99\%$ region in increasing lightness respectively.}
    \label{fig:energ_resids}
\end{figure}

In Fig. \ref{fig:energ_resids}, we show the distribution of percentage residual differences between \texttt{RTFAST} and \texttt{RTDIST} (see eq. \ref{eq:resid} $\times 100\%$) as a function of energy for the same test set. This representation averages over the entire parameter space and instead shows general trends in errors for the emulator. The slight increasing of percentage error difference towards the highest and lowest energies is loosely related to the increasing steepness of the power law slope with higher $\Gamma$. As the most significant changes from the continuum (absorption edges and emission lines) do not deviate significantly in error from the errors in the rest of the continuum, \texttt{RTFAST} appears to capture much of the non-linear behaviour resultant from the general relativistic effects that effect the reflection spectrum. This partly shows the effectiveness of utilising PCA decomposition to preserve correlation between energy bins by ensuring that the continuum does not dominate the spectrum in training and neglect the emulation of emission lines due to their small contribution to the overall spectrum.

\subsection{Recovering parameters} \label{fit}

In this section, we show that \texttt{RTFAST} can retrieve posteriors for simulated data from \texttt{RTDIST}. As the original model contains multiple degenerate pairs of parameters (notably degeneracy between the distance and mass of the black hole in the inferred ionisation of the disk in Eq. 10 of I22) and some of the parameters cannot be constrained from the spectra alone, we constrain the problem to a smaller suite of parameters in which we can retrieve reasonable posteriors.

\begin{table}
    \centering
    \resizebox{\linewidth}{!}{%
    \begin{tabular}{l|c|l}
        Parameter & Description & Prior \\ \hline
        $h$ ($R_g$)  & \begin{tabular}[tl]{@{}c@{}}height of the corona in black hole \\ gravitational radii\end{tabular} &  $\log\mathcal{U}(1.5,10)$ \\ 
        $a$   & spin &  $\mathcal{U}(0.5,0.998)$   \\ 
        $i$ ($^\circ$)   & inclination (degrees) &  $\log\mathcal{U}(30,80)$  \\ 
        $r_{\rm in}$ (ISCOs)   & inner radius of disk &  $\log\mathcal{U}(1,20)$  \\ 
        $\Gamma$  & photon index&  $\mathcal{U}(2,2.75)$  \\ 
        $D$ ($10^5$kpc) & Distance in kpc &  $\log\mathcal{U}(3.5,500)$  \\ 
        $A_{\rm fe}$ & Iron abundance in solar units&  $\log\mathcal{U}(0.5,3)$ \\ 
        log($n_e /{\rm cm}^{-3}$)  & Electron density of the disk&  $\mathcal{U}(15,20)$ \\ 
        $N_h$ ($10^{20}{\rm cm}^{-2}$) & Hydrogen column density& $\log\mathcal{U}(10^{-3},1)$ \\ 
        $A_{\text{norm}}$ ($10^{-4}$) & \texttt{RTDIST} normalisation & $\log\mathcal{U}(10^{-4},0.1)$ \\ 
    \end{tabular}}
    \caption{Priors for a more realistic fitting scenario. $\mathcal{U}(a,b)$ and $\log\mathcal{U}(a,b)$ indicate uniform and log uniform priors bounded between lower bound $a$ and upper bound $b$.}
    \label{table:priors}
\end{table}

\begin{table}
    \centering
    \begin{tabular}{|l|l|l|}
        Parameter               & Input         & Fitted parameters         \\ \hline
        $h$ ($R_g$)             & 6             & $4.81^{+0.60}_{-0.48}$    \\ 
        $a$                     & 0.9           & $0.64^{+0.05}_{-0.06}$    \\ 
        $i$ ($^\circ$)          & 57            & $57.78^{+0.40}_{-0.42}$   \\ 
        $r_{\rm in}$ (ISCOs)    & 1             & 1                         \\ 
        $r_{\rm out}$ ($R_g$)   & $2\times10^4$ & $2\times10^4$             \\
        $z$                     & 0.024917      & 0.024917                  \\
        $\Gamma$                & 2.45          & $2.44^{+0.01}_{-0.01}$    \\ 
        $D$ ($10^5$ kpc)        & 1             & $0.77^{+0.087}_{-0.073}$  \\ 
        $A_{\rm fe}$            & 1             & $0.98^{+0.03}_{-0.03}$    \\ 
        log($n_e /{\rm cm}^{-3}$)  & 17            & $16.41^{+0.08}_{-0.07}$   \\ 
        $kT_e$                  & 50            & 50                        \\
        $\frac{1}{\beta}$       & 1             & 1                         \\
        $M_{BH}$                & $3\times10^6$ & $3\times10^6$             \\
        $h/r$                   & 0.02          & 0.02                      \\ 
        $b_1$                   & 0             & 0                         \\
        $b_2$                   & 0             & 0                         \\ 
        $N_h$ ($10^{20}{\rm cm}^{-2}$)   & 5             & $4.82_{-0.05}^{+0.05}$    \\ 
        A  ($10^{-4}$)          & 2.2           & $3.1^{+0.6}_{-0.3}$       \\ 
    \end{tabular}
    \caption{Input parameters and posterior fitted parameters for the simulated observations with $3\sigma$ uncertainties as recovered from fitting with \texttt{RTFAST}.}
    \label{table:fits}
\end{table}

We simulate an observation with 260ks XMM-Newton exposure with \texttt{xspec}'s \citep{arnaud1996xspec} fakeit routine, basing our source parameters on Ark 564 \citep{kara2013discovery,kara2016global,ingram2022}, listed in Table \ref{table:fits}. We use the open source software \texttt{nDspec} (\url{https://github.com/matteolucchini1/nDspec}, Lucchini et al (in prep)) for implementation of instrument response convolution throughout the rest of this paper. We perform Bayesian parameter estimation using the \texttt{emcee} package \citep{foreman2013emcee}.  

We specify our priors in Table \ref{table:priors}, but fix all parameters to their true values other than $h$, $a$, $i$,$\Gamma$, $D_{\text{kpc}}$, $A_{\text{fe}}$, $\log{N_e}$, $N_h$, and $A_{\text{norm}}$. We fix $z$ as redshift constraints for AGN can be found from observations in other wavelengths. For simplicity, we fix inner radius to the ISCO as x-ray reflection models are generally sensitive to either spin or inner radius. We fix $h/r$ (scale height of the disk), $b_1$ and $b_2$ as these have been generally unconstrained in previous fits to data with \texttt{RTDIST}. We fix $kT_e$ as we are unable to observe the high energy roll-over at approximately $50 \text{keV}$ due to XMM-Newton only observing between $0.1$ and $20 \text{keV}$.

Due to the exceptionally high counts and to speed up computations, we use a Gaussian likelihood. Errors are generated by \texttt{xspec}'s fakeit routine that incorporate the XMM-Newton PN background spectrum as well as Poissonian uncertainties. We use the spectrum between $0.3$ keV and $10$ keV. We employ 100 walkers with 70000 steps each for our MCMC, and otherwise use the default parameters.

\begin{figure}
    \centering
    \includegraphics[width=0.99\linewidth]{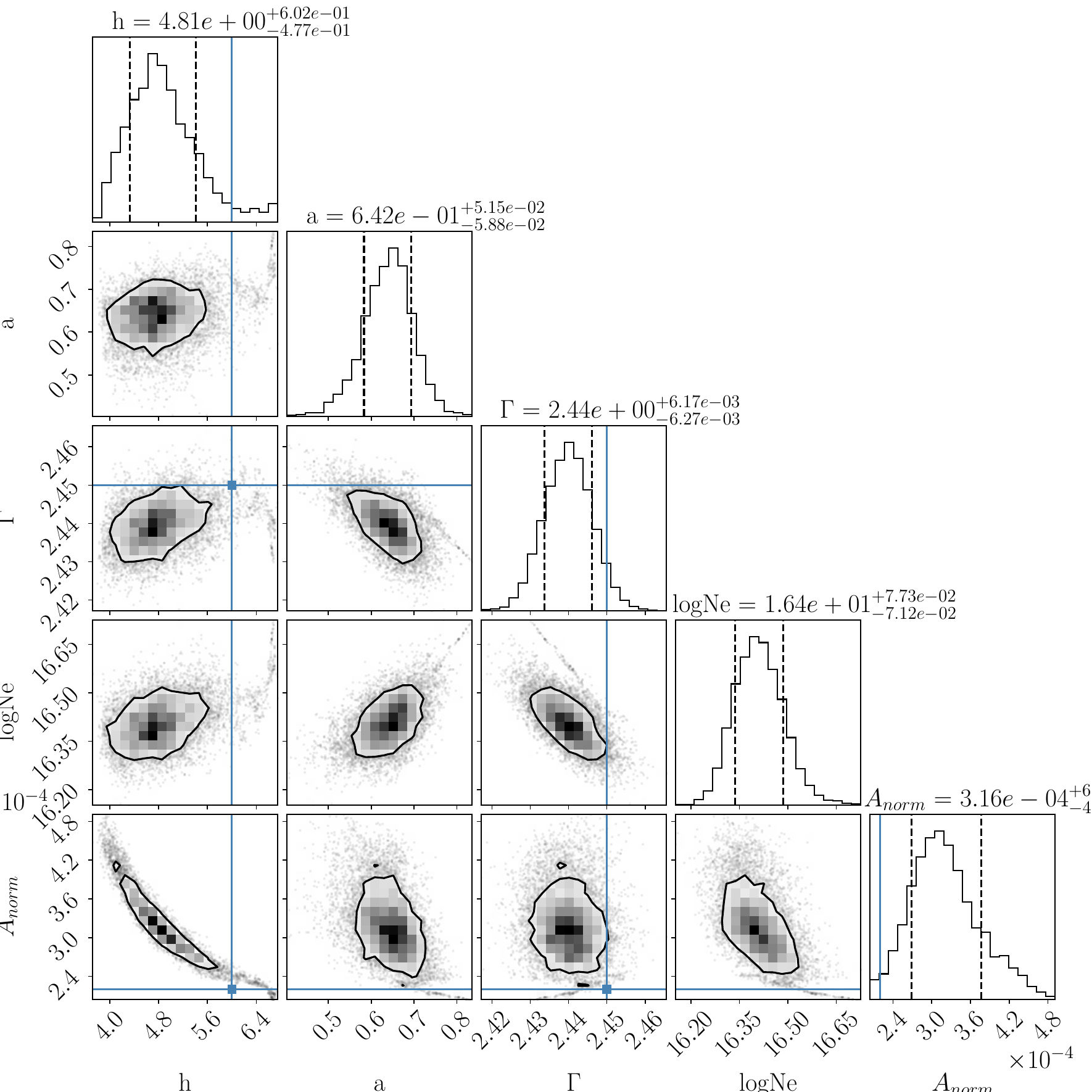}
    \caption{Posteriors of fitting for a more realistic scenario where 2 sigma contours are plotted. \myRed{The plot limits indicate the 3 sigma boundary.} We show 5 of the 8 parameters varied. The full posteriors can be found in appendix \ref{full_post}. The blue lines show the true parameters used to generate the simulated observation.}
    \label{fig:fit}
\end{figure}

Fig \ref{fig:fit} shows the posteriors for 5 parameters: height, spin, $\Gamma$, $\log N_e$ and $A_{\text{norm}}$ (a corner plot showing the posteriors of all the free parameters is presented in Appendix \ref{full_post}). Table \ref{table:fits} shows the full list of constrained parameters. The 1-dimensional posteriors \myRed{for the majority of parameters} include the truth within the 3 sigma bounds, and sometimes fall within the 2 sigma bounds of the posterior. The two-dimensional posteriors show a clear bias in the some of the inferred parameters and correlations between parameters.  Notably, the relationship between $A_{\text{norm}}$, a proxy for the accretion rate, and $h$, the height of the corona, shows a clear "banana" degeneracy. This is realistic since a higher accretion rate is required to create the necessary reflection observed in the spectrum. There is also a bias towards lower values of spin and $\log N_e$ for this fit \myRed{, leading the true value to lie outside of the 3 sigma bounds of the posterior.}

\begin{figure}
    \centering
    \includegraphics[width=0.99\linewidth]{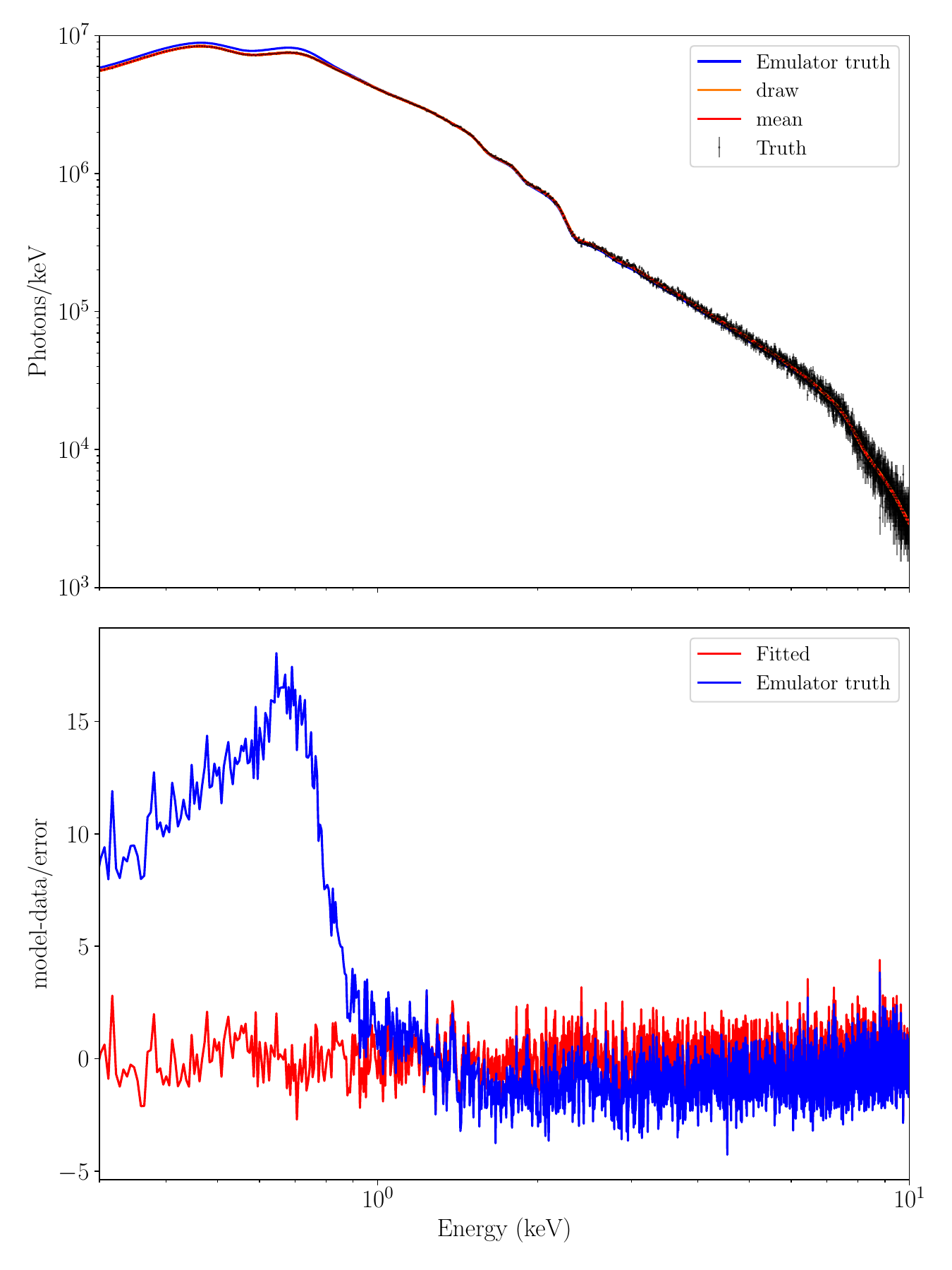}
    \caption{Plots of the fitted spectrum, the emulated spectrum when the true parameters are inputted and the true spectrum generated by \texttt{RTDIST}. Posterior draws are plotted in orange, while the mean of these draws is plotted in red. The true simulated spectrum is plotted in black, with errors from Poisson noise and systematics. The emulator output for the true parameters is plotted in blue.}
    \label{fig:draws}
\end{figure}

This bias in some recovered parameters is physically explainable. Fig \ref{fig:draws}, top panel, shows an observation simulated with \texttt{RTDIST} along with the spectrum predicted by the emulator for the exact same set of input parameters. We also show example spectra generated from posterior draws, and the mean of those posterior draws. The bottom panel shows the error-weighted residuals of the mean of the posterior draws alongside the same error-weighted residuals of the "correct" spectrum given the true parameters predicted by the emulator. The \texttt{RTFAST} spectrum over-predicts a considerable soft excess. As a result, the model will underestimate $\log N_e$ to compensate for this bias during inference. 

\begin{figure}
    \centering
    \includegraphics[width=0.99\linewidth]{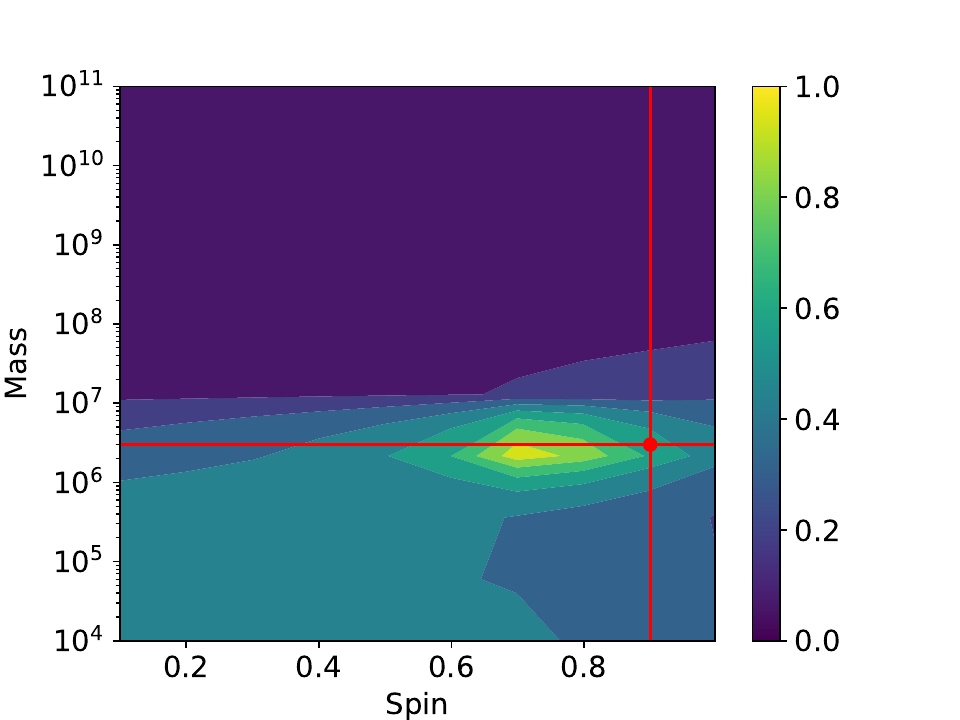}
    \caption{The two dimensional log-likelihoods of fitting the spectrum, varying mass and spin, plotted in contours through parameter space. Brighter contours indicate higher likelihood. \myRed{The red lines indicate the true parameters and what should be the peak of likelihood is represented by the red dot where they intersect.}}
    \label{fig:2dlikelis}
\end{figure}

\begin{figure}
    \centering
    \includegraphics[width=0.99\linewidth]{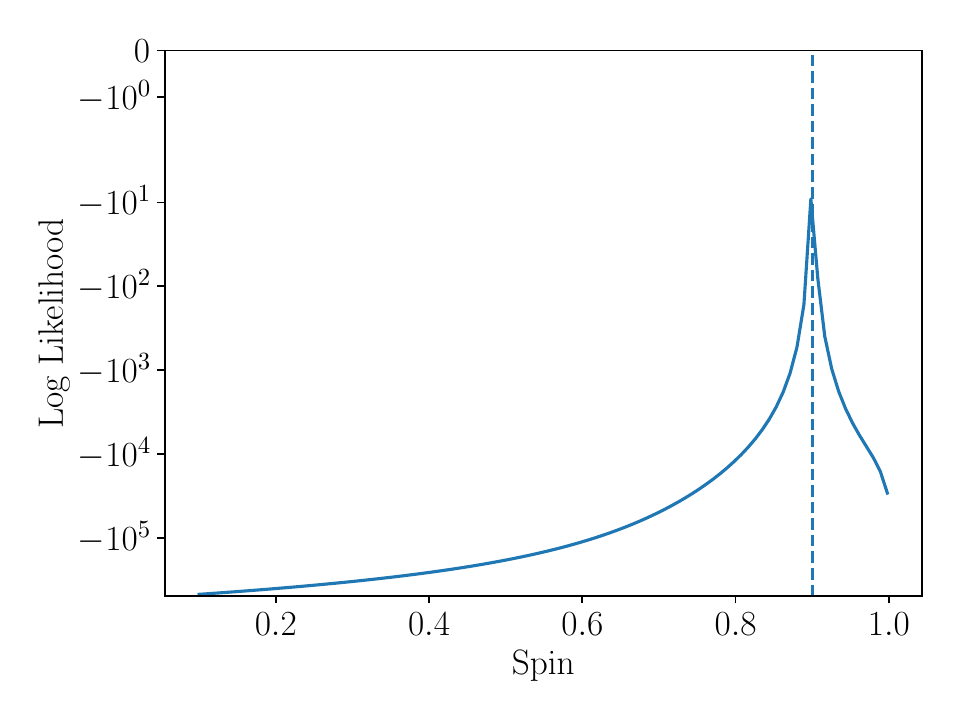}
    \caption{The log likelihood as a function of spin when all other parameters are kept fixed. The true spin is plotted as a vertical dashed line at 0.9}
    \label{fig:spin_likeli}
\end{figure}

The inference of lower spin is better understood when freezing all parameters to the true values and then evaluating the likelihood on a 2-d grid of mass and spin. In Fig \ref{fig:2dlikelis}, we took a grid of 10 points in mass and spin and evaluated the log-likelihood of each point on the grid. We took the normalized logarithm of the negative log-likelihood and plotted the contours. For a specified mass of $3\times10^6 M_{\odot}$, the spin with the highest likelihood is around $0.7$. In contrast, Fig. \ref{fig:spin_likeli} plots the likelihood as a function of spin when all other parameters are fixed to the true value. The point of maximum likelihood is clearly located extremely close to the truth, identifying a high, but not maximal spin as the most likely outcome. This implies a bias in the emulator such that it produces a spectrum that would be identified at lower spin at a slightly smaller mass for the original numerical model. When mass is fixed, the effects of spin on the spectrum are well-modelled and results in the correct inferred spin. 

We plot the 2-dimensional likelihoods in the same manner in Appendix \ref{likelis} for a number of other parameters combinations which show this bias in the emulator predominately drives a bias in the inferred mass spin likelihood for this particular simulated data. We also plot similar figures as Fig. \ref{fig:spin_likeli} for all parameters in Appendix \ref{likelis}.

This significant bias in the modelling is partly caused by the emulator's unusually bad performance for this particular set of parameters of Ark 564. This particular spectrum has deviations of more than $10\%$ in the soft X-ray which places it within the $5\%$ worst performing spectra for the emulator. This amount of bias in inferred parameters is thus an upper limit to the emulator's worst performance. The bias in the model could be mitigated in this instance by fitting the spectrum between $1$ and $10$ keV, as the model is otherwise accurate for the rest of the spectrum. Notably, modelling the soft excess in the x-ray below 1-2keV has been a well-known problem for x-ray reflection and often requires additional model components \citep{lewin2022x,masterson2022evolution,xu2022ejection}. Using \texttt{RTFAST} between $1$ and $10$ keV more generally may be a better approach to mitigate biases caused by the emulator (particularly as the emulator performs the worst between $0.1-1\text{keV}$ and $10-20\text{keV}$ (which can be seen in Fig. \ref{fig:energ_resids}) as well as avoid issues in modelling of the soft excess. 

Modelling actual data is much more difficult. Maximum likelihood fits to the data may be unstable, and one may achieve different best fits for different starting parameters, indicating that degeneracies are present in the data. Thus, it often requires making assumptions about the geometry of the system to eliminate degeneracies by fixing certain parameters, similar to other analyses with \texttt{RTDIST} when fitting to real data \citep{nathan2024proof}. Including timing products alongside the spectrum in the inference assists greatly with reducing degeneracy (as seen in I22) but cannot solve all of these challenges. \texttt{RTFAST} will allow for greater exploration of these degeneracies by enabling fast posterior calculations, even for complex physical models.

\subsection{Poor extrapolation} \label{outofbounds}

\begin{figure}
    \centering
    \includegraphics[width=0.99\linewidth]{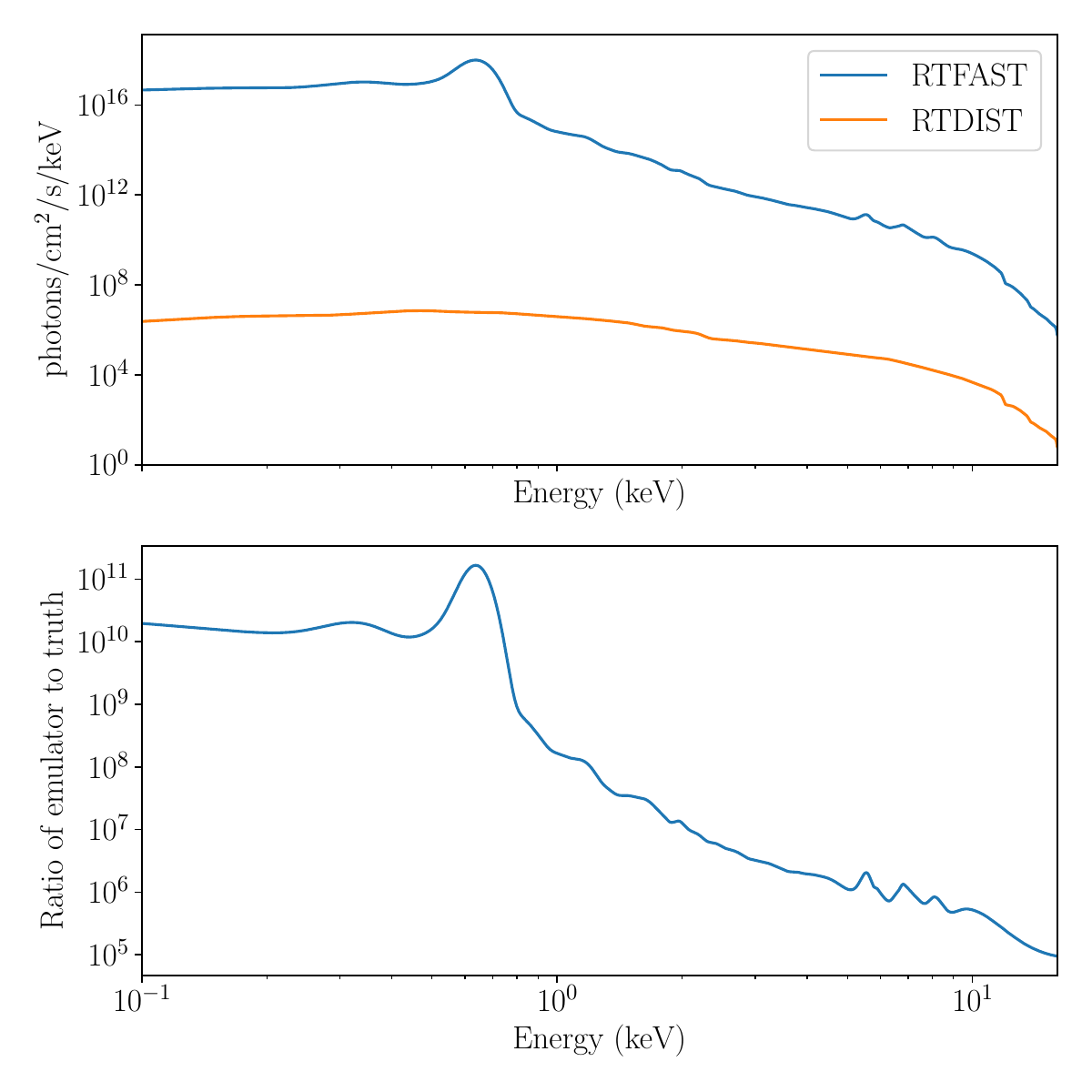}
    \caption{Spectrum generated by \texttt{RTFAST} outside of the training bounds compared to \texttt{RTDIST}. We set coronal height ($h$) to $400 R_g$ and we observe that \texttt{RTFAST} produces extremely unrealistic photon fluxes. This is a clear demonstration of \texttt{RTFAST} being unable to extrapolate behaviour.}
    \label{fig:ooobounds}
\end{figure}

In this section, we show the pitfalls of calling \texttt{RTFAST} outside of the training data bounds. Neural networks and other machine learning techniques have been repeatedly shown to extrapolate extremely poorly outside of their training bounds \citep{pastore2021extrapolating} and \texttt{RTFAST} is not an exception to this. In Fig. \ref{fig:ooobounds}, we generate a sample with a coronal height, $h$, of 400 $R_g$ from both \texttt{RTFAST} and \texttt{RTDIST}. We can clearly see that \texttt{RTFAST} gives a completely nonsense result, overestimating the spectrum by many orders of magnitude. Note that for other parameters than $h$, the effect can be more subtle and is more likely to be misleading. Users should never, under any circumstance, utilize parameter bounds outside those imposed in section \ref{phys_prior}.

\section{Discussion and conclusions} \label{conclusions}

\subsection{Emulator}

We have designed and implemented \texttt{RTFAST}, an emulator that acts as a drop-in replacement for the model \texttt{RTDIST} for AGN. \texttt{RTFAST} is able to recreate theoretical model spectra to within $3\%$ precision of the original model with an $\mathcal{O}(10^2)$ speed up in conservative conditions. This is the first emulator for the \texttt{reltrans} model suite and the first emulator to achieve this level of precision for this many physically meaningful model parameters within X-ray spectroscopy. The wide range of magnitude in model spectra has previously made this difficult to achieve when training neural networks to this level of precision. 

By enabling computations on the GPU, the emulator allows us to harness parallelization to yield an $8000$ times speed up over the original model when called sequentially. Because neural networks are differentiable models, where we can compute gradients of the model output as a function of input parameters, the emulator also allows the use of modern optimization and sampling algorithms to greatly speed up analysis. This in turn enables full full Bayesian analysis on AGN and, in future, X-ray binaries. We will use this in future work to perform more robust measurements of black hole properties and perform simultaneous modelling of multiple sources to retrieve information about the black hole population as a whole. 

Upon extensive testing in section \ref{Robust}, we have found that there are small biases to inferred parameters introduced by \texttt{RTFAST}. These small biases are physically explainable and the emulator seems to capture important physical relationships between parameters. The biases depends on the section of parameter space being investigated, as well as what selection of parameters are being inferred, and any users of \texttt{RTFAST} should be careful about where these biases can creep into their interpretation of an observation. We suggest that users should always check the fitted model and plot posterior draws from the original \texttt{RTDIST} to assess whether they should trust the final results coming from RTFAST. 

It should also be noted that the computational gains of \texttt{RTFAST} are relatively independent of the inclusion of additional physics in the case of \texttt{RTDIST}. Introducing complexity into the physical model is likely to introduce only small differences to the final model output, thus the complexity of \texttt{RTFAST} will likely have to increase only marginally to be able to model these scenarios. Even if future iterations of \texttt{RTDIST} were to take 100 seconds to evaluate instead of $\approx 0.5$ seconds, \texttt{RTFAST} will still take on order of $10^{-3}$ seconds to evaluate. We suggest that emulators such as \texttt{RTFAST} will allow us to probe far more complex physical scenarios while still being able to perform the millions of evaluations required for proper inference of the posteriors.

For those in the community who wish to implement this type of work on their own models, we recommend reducing the complexity and size of the model's output (here: the spectrum) using algorithms that preserve important relationships in the model (e.g.~PCA). For instance, for models that feature large suites of absorption and emission lines, preserving their appearance as a function of input parameters in a way that the neural network does not need to learn will allow developers to keep the network size low. Domain knowledge of both the model and astronomical sources modelled was key in constraining the problem as highlighted in Section \ref{phys_prior}. When developing emulators, we recommend considering both observational and theoretical constraints to the model parameter space and the model itself. We discuss this in more detail in Appendix \ref{build_emus}.

\subsection{Bayesian analysis}
In section \ref{speed}, we showed that \texttt{RTFAST} is, at its most conservative estimation, 200 times faster than the original \texttt{RTDIST} model when called sequentially. This becomes even faster once considering the ability to evaluate multiple models in vector form---calculating spectra up to $\mathcal{O}(10^4)$ times faster than the original model. Current MCMC inference using \texttt{RTDIST} can take up to a month or more, particularly if multiple datasets are fit jointly. Using \texttt{RTFAST} reduces those fits to a few hours at most. This allows for further in-depth exploration of the posteriors of the degenerate parameter space of \texttt{RTDIST}, illustrated in section \ref{fit}. 

\texttt{RTFAST} was developed to make it possible to properly explore the large parameter space of \texttt{RTDIST}. \texttt{RTFAST} enables us to do this via multiple avenues: Monte-Carlo Markov-Chain (which already has widespread use in the field), nested sampling (which has already been somewhat adopted by some of the community; \citealt{buchner2014x}), Hamilton Monte Carlo (a Monte Carlo variant that uses gradients in respect to parameters to achieve faster optimization; \citealt{DUANE1987216}) and simulation based inference (a machine learning technique that approximates the posterior of an observation via simulations of a theoretical model; \citealt{cranmer2020frontier}). All of these techniques (with the exception of Hamilton Monte Carlo) are already possible to use with \texttt{RTDIST} but are prohibitively computationally expensive to use practically. We have implemented all of these techniques in an accompanying notebook to \texttt{RTFAST} in the RTFAST repository \url{https://github.com/SRON-API-DataMagic/RTFAST}.

It is possible to use \texttt{RTFAST} to produce MCMC chains to a considerable length towards convergence and then switch to \texttt{RTDIST} for the final steps in the chain towards convergence to avoid the biases potentially present in sampling posteriors with \texttt{RTFAST}. This allows for a dramatic reduction in computation time while still retrieving the complex posteriors.

\subsection{Future work} \label{future}
\texttt{RTFAST} approximates the spectrum generated by the \texttt{RTDIST} model, but \texttt{RTDIST} comprises a number of different physical processes (i.e. reflection from the disk \citep{garcia2013x}, relativistic effects on photons emitted by the corona \citep{dauser2016relativistic}, etc). Our next-generation emulator will focus on reproducing these individual effects using the design philosophy of \texttt{RTFAST} as a template. This would allow for a modular design that enables easy enabling and re-enabling of the true underlying physical model. An example of this would be to emulate the disk illumination profile resulting from the geometry of the system rather than computing it directly with ray-tracing. Retraining a single component may also result in less required training data when updating the emulator compared to retraining the emulator on the total output.

\texttt{RTFAST} currently contains systematic errors of order of $3\%$ which is just accurate enough to be comparable with the systematic errors of current generation X-ray instruments such as XMM-Newton. We would like to achieve better precision in recreating the spectrum in the future, especially to reduce or even eliminate bias in the inferred parameters in the future. We expect to be able to reduce the error by at least one order of magnitude by changing our strategy slightly. 

Currently, we emulate all components in the spectrum at once. The \texttt{RTDIST} spectrum is mainly made up of 2 components: a comptonized power law approximating the corona (\texttt{nthcomp}), and a reflection spectrum resulting from reprocessed coronal emission from the disk. The reflection spectrum itself is by far the most computationally expensive component of \texttt{RTDIST} as it requires the calculation of relativistic effects. This component typically composes approximately $10\%$ of the total flux in observed spectra from both AGN and X-ray binary sources. Emulating the reflection spectrum alone would mean that the resultant error in the total spectrum would be an order of magnitude smaller and prevent strong biases in parameters inferred from the continuum. This would likely require returning to the \texttt{reltransDCp} flavour of \texttt{reltrans}, due to the entanglement of normalisation and the spectrum by $A_{\text{norm}}$ in \texttt{RTDIST}.

Another potential extension would be to emulate the model after convolution with the instrument response. This would allow for the bottleneck of convolution of the instrument response to be avoided entirely, a topic that is particularly important for instruments such as XRISM which feature extremely large, high resolution instrument responses. We opt to not adopt this strategy in this paper to keep the emulator somewhat instrument independent.

\texttt{RTDIST} contains functionality to compute timing products caused by reverberation between the corona and disk. In the future, we will design an instrument-independent emulator that will allow for the computation of \texttt{RTDIST} timing products to further constrain black hole parameters. This is outside of the scope of this paper due to the complexity of emulating the cross-spectrum, caused by its higher dimensionality and greater range in model outputs that could not be constrained in the same way as in section \ref{phys_prior}.

While \texttt{RTFAST} is an ensemble where the final outputs of its constituent neural networks are simply averaged together, there are potentially better approaches to this ensemble technique such as the weighting of models by their relative accuracy to the final spectrum or explicitly training a considerably larger ensemble and pruning networks to find a more optimal combination of networks to better approximate \texttt{RTDIST}.

During development, we attempted use of an active learning technique called Query by Dropout Committee \citep{ducoffe2015qbdc}. This technique attempts to select the most effective data to generate during training to reduce the total amount of data required to train \texttt{RTFAST}. This ultimately was not used in \texttt{RTFAST} as generating data during training became unwieldy when performing architecture and learning sweeps in section \ref{architecture} and \ref{training}. This strategy will likely be more appropriate for the cross-spectrum as it both takes more time to compute and will be larger in data-size than the spectrum alone. As such, we would like to re-attempt this technique when emulating the cross-spectrum. Another potential technique for reducing the amount of required training data is active subspace sampling \citep{constantine2014active}. Active subspace sampling uses evaluations of the gradient to find the areas of strongest variability, and subsequently uses these directions to construct a sampling surface on a lower-dimensional subspace.

\section*{Acknowledgements}

We would like to thank the referee for their constructive comments. This work used the Dutch national e-infrastructure with the support of the SURF Cooperative using grant no. EINF-9197 and EINF-5505. We'd like to express our thanks to D.Burke in particular and the CIAO help desk for their assistance in the use of the sherpa package within this paper. AI acknowledges support from the Royal Society.

\section*{Data Availability}

The training data generated for this project is too large to be open access, however the parameters for all generated spectra in this project can be found at \url{https://doi.org/10.5281/zenodo.14264681}. The RTDIST model used to generate the training data is currently not available publicly, but is available on request. The training program is publicly available and can be found at \url{https://github.com/SRON-API-DataMagic/rtdist-emulator}. If a user wishes to use the trained emulator for fitting of observations, they can find the trained model and the required program to install and use it at \url{https://github.com/SRON-API-DataMagic/RTFAST}.



\bibliographystyle{mnras}
\bibliography{biblio} 




\appendix

\section{Physical prior distributions}\label{full_dist}

\begin{figure*}
    \centering
    \includegraphics[width=0.99\linewidth]{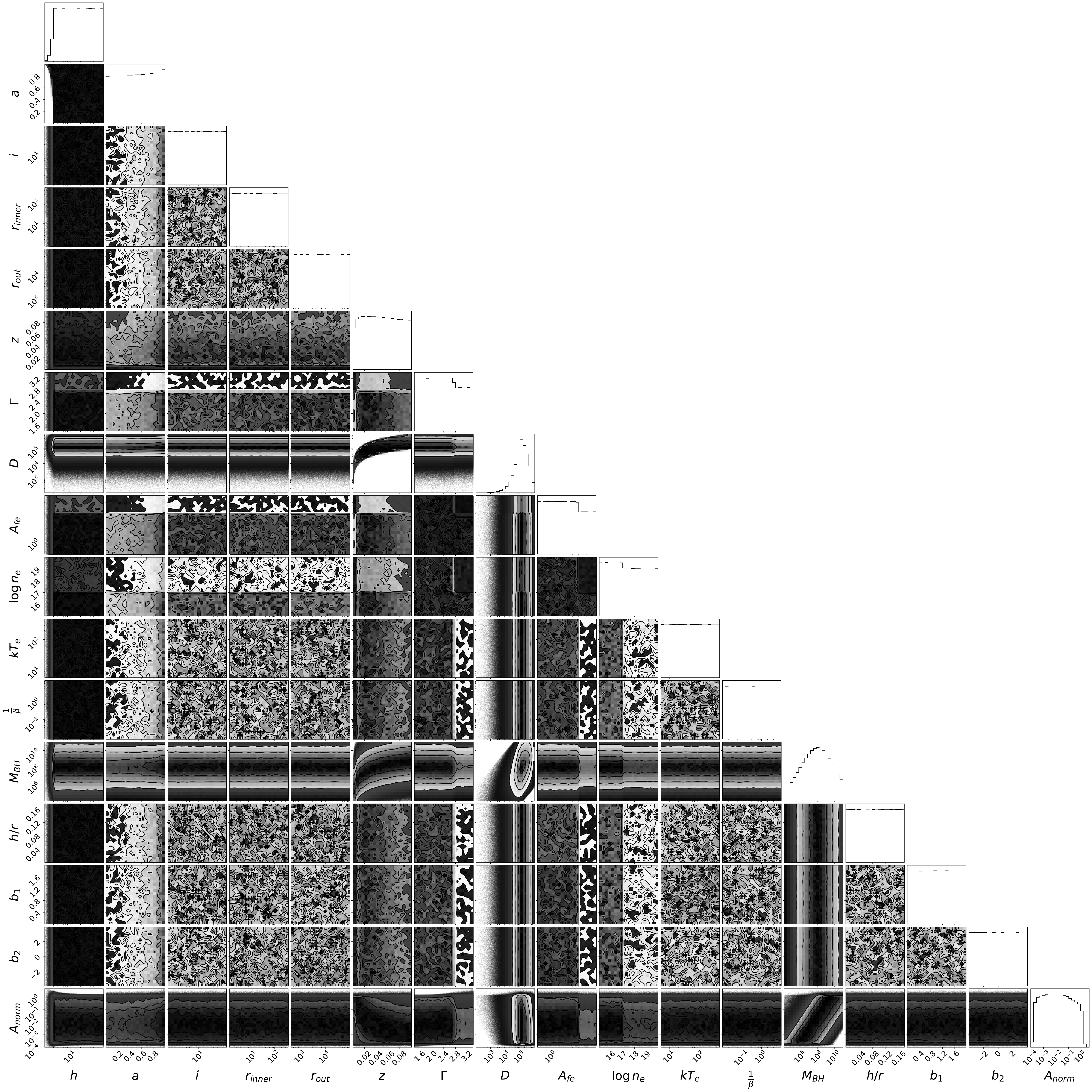}
    \caption{Corner plot of all training set parameters and their distributions. Possible training data to be generated is constrained by the physical constraints outlined in section \ref{phys_prior}.}
    \label{fig:full_phys_prior}
\end{figure*}

In Fig \ref{fig:full_phys_prior}, we plot the distributions of all training parameters as specified in section \ref{phys_prior}.

\section{Full posteriors}\label{full_post}
\begin{figure*}
    \centering
    \includegraphics[width=0.9\linewidth]{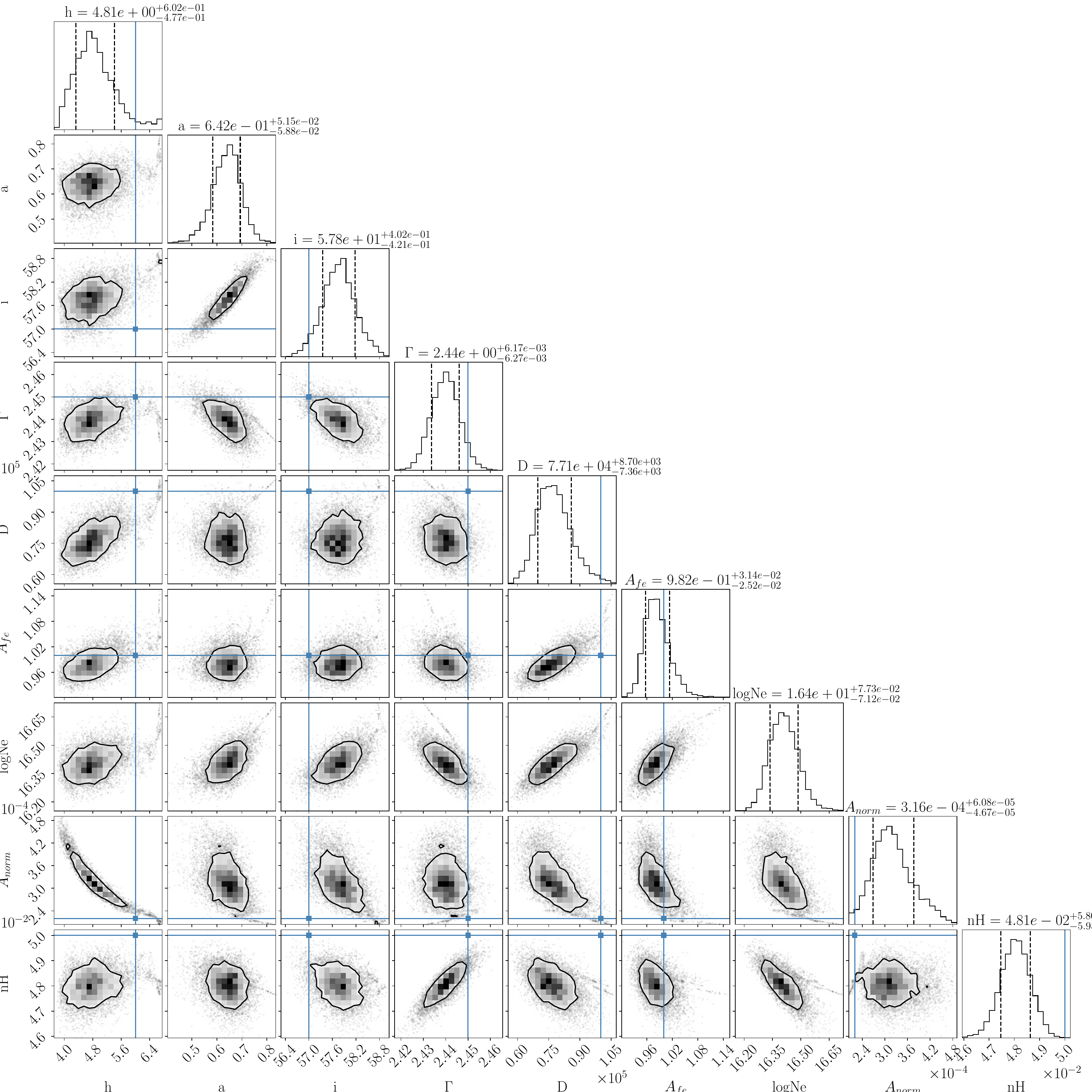}
    \caption{Posteriors of fitting. The 2-sigma contours are explicitly plotted. The blue lines show the true parameters used to generate the simulated observation.}
    \label{fig:fit_full}
\end{figure*}

\begin{figure}
    \centering
    \includegraphics[height=0.8\paperheight]{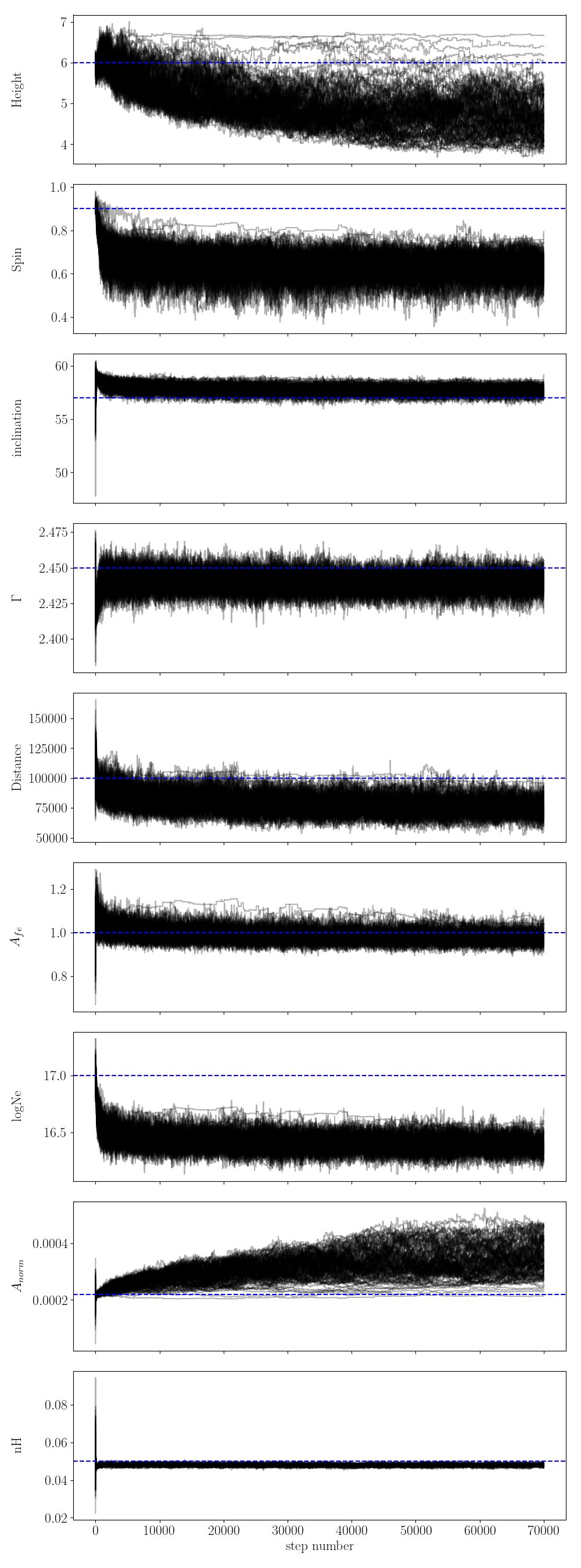}
    \caption{MCMC chains as a function of steps of each walker. The blue dotted line indicates the true parameters.}
    \label{fig:chain}
\end{figure}

In Fig \ref{fig:fit_full}, we report the full posteriors of the fit in section \ref{fit}. We also plot the MCMC chains in their entirety in Fig \ref{fig:chain}.

\section{Likelihood space} \label{likelis}

\begin{figure}
    \centering
    \includegraphics[width=0.99\linewidth]{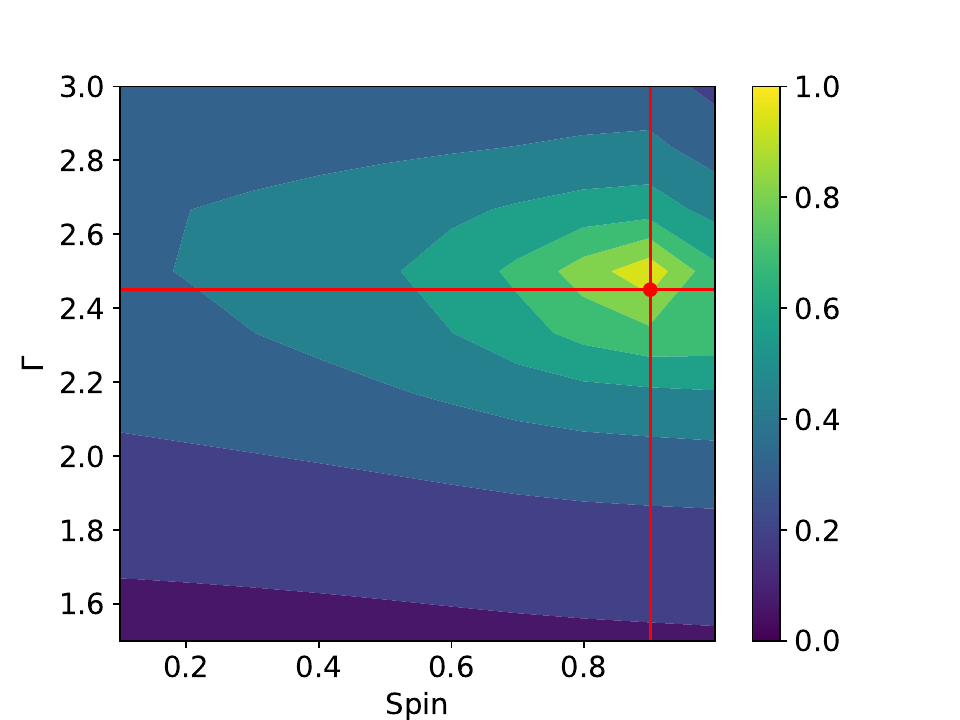}
    \caption{2-dimensional grid of the likelihood for spin vs $\Gamma$.}
    \label{fig:spin_gam}
\end{figure}

\begin{figure}
    \centering
    \includegraphics[width=0.99\linewidth]{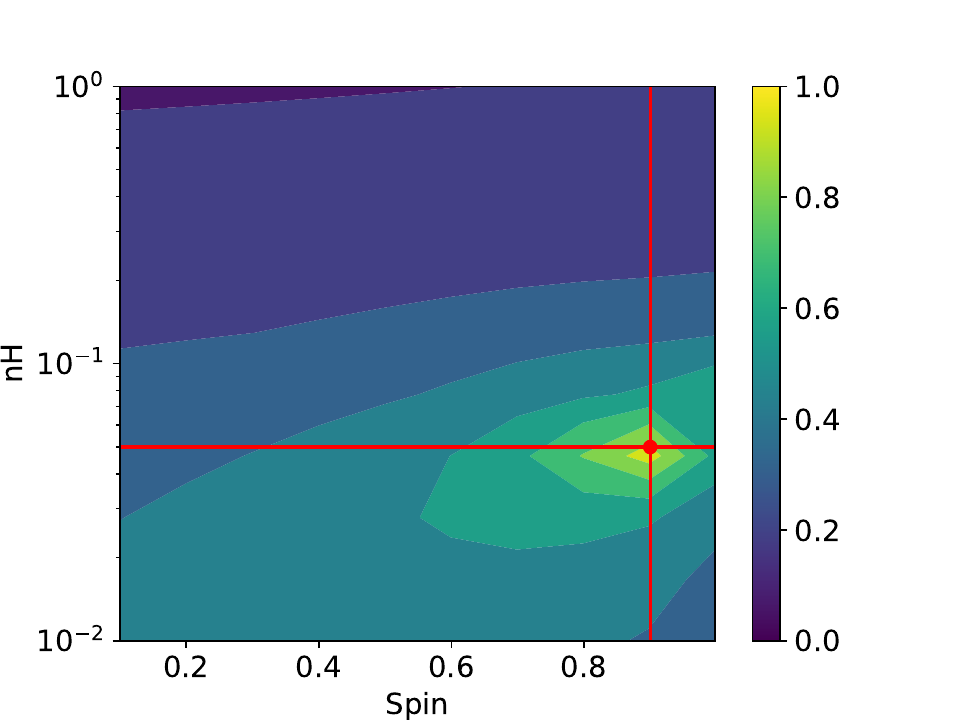}
    \caption{2-dimensional grid of the likelihood for spin vs $n_H$.}
    \label{fig:spin_nh}
\end{figure}

\begin{figure}
    \centering
    \includegraphics[width=0.99\linewidth]{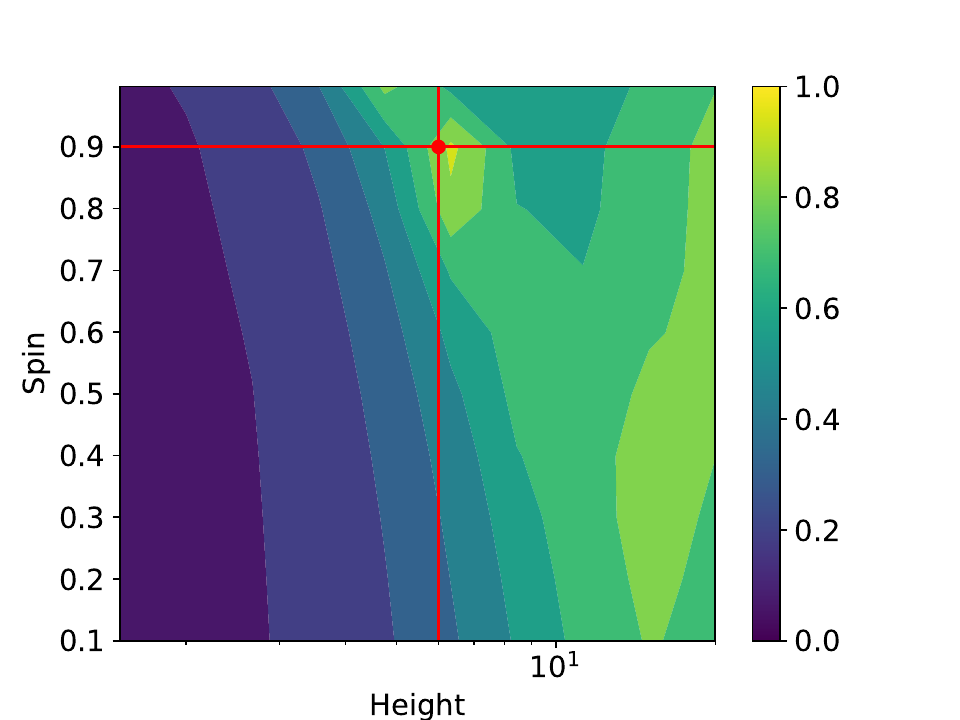}
    \caption{2-dimensional grid of the likelihood for coronal height vs spin.}
    \label{fig:h_spin}
\end{figure}

\begin{figure}
    \centering
    \includegraphics[width=0.99\linewidth]{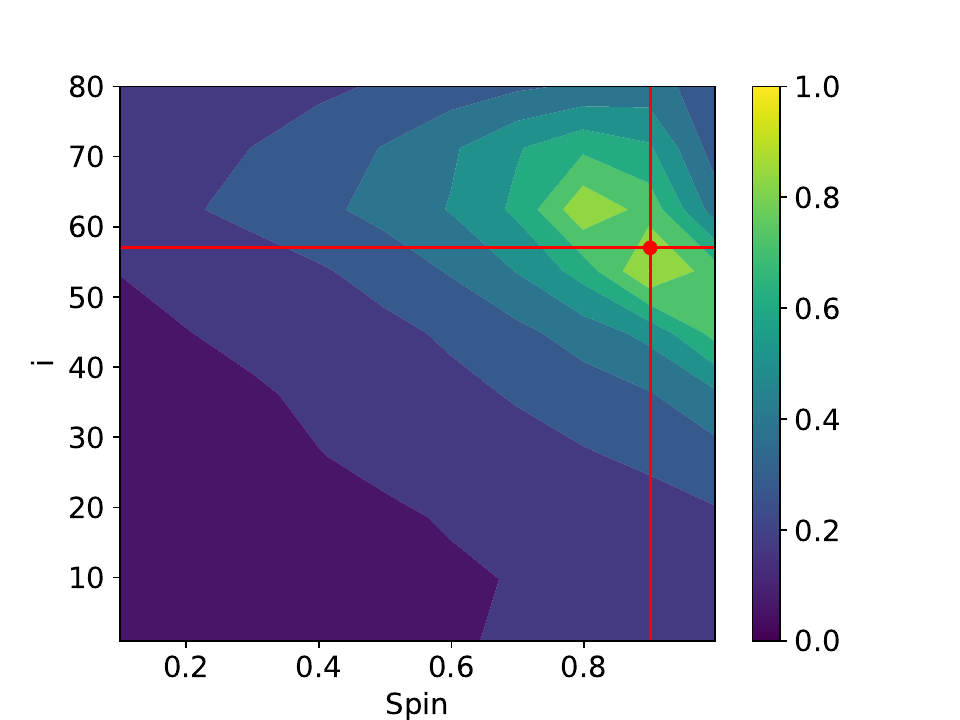}
    \caption{2-dimensional grid of the likelihood for spin vs inclination.}
    \label{fig:spin_i}
\end{figure}

\begin{figure}
    \centering
    \includegraphics[width=0.99\linewidth]{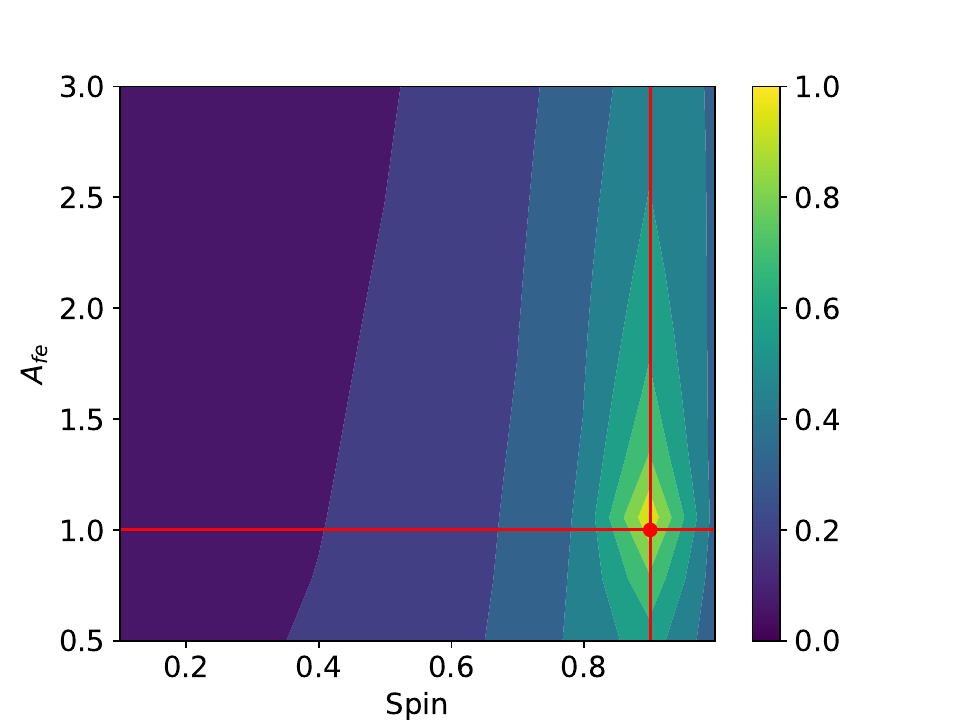}
    \caption{2-dimensional grid of the likelihood for spin vs $A_{fe}$.}
    \label{fig:spin_Afe}
\end{figure}

\begin{figure}
    \centering
    \includegraphics[width=0.99\linewidth]{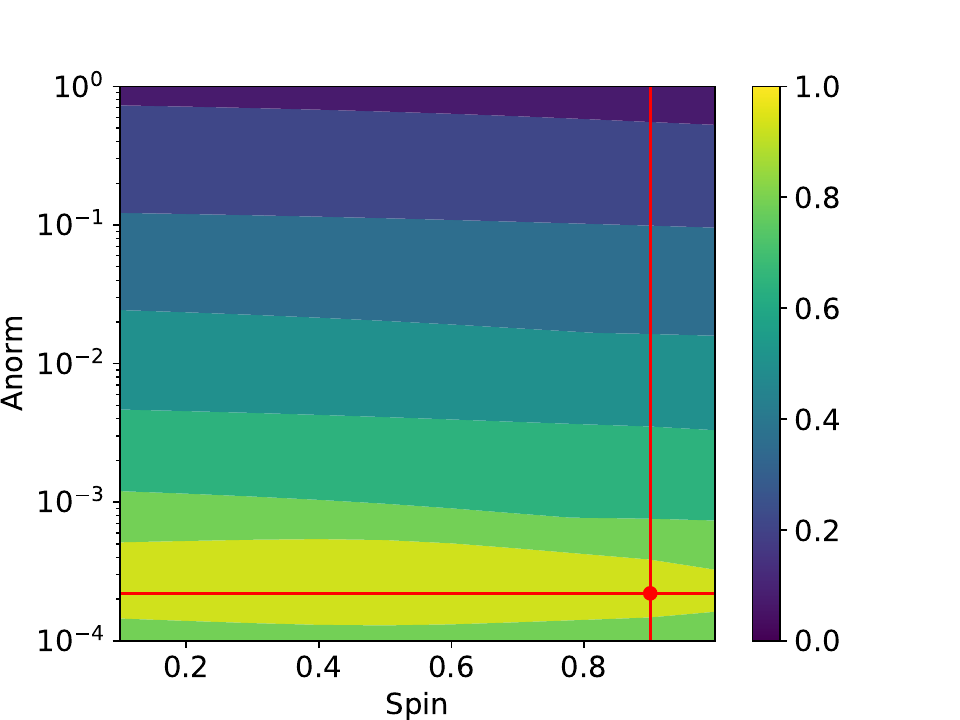}
    \caption{2-dimensional grid of the likelihood for spin vs $A_{norm}$.}
    \label{fig:spin_anorm}
\end{figure}

\begin{figure}
    \centering
    \includegraphics[width=0.99\linewidth]{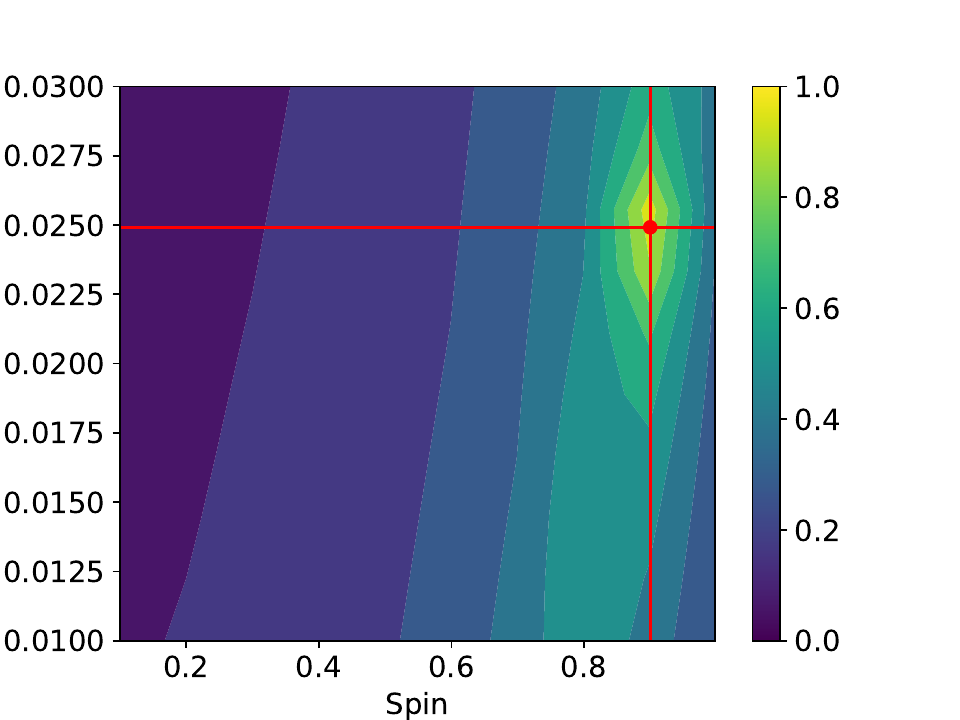}
    \caption{2-dimensional grid of the likelihood for spin vs z.}
    \label{fig:spin_z}
\end{figure}

\begin{figure}
    \centering
    \includegraphics[width=0.99\linewidth]{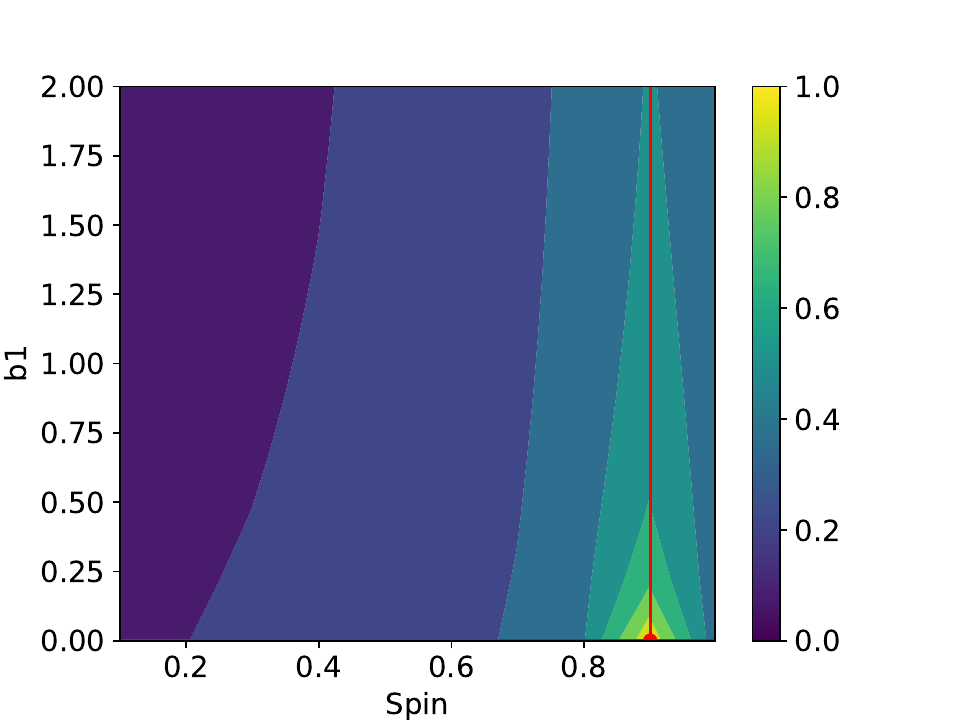}
    \caption{2-dimensional grid of the likelihood for spin vs b1.}
    \label{fig:spin_b1}
\end{figure}

\begin{figure}
    \centering
    \includegraphics[width=0.99\linewidth]{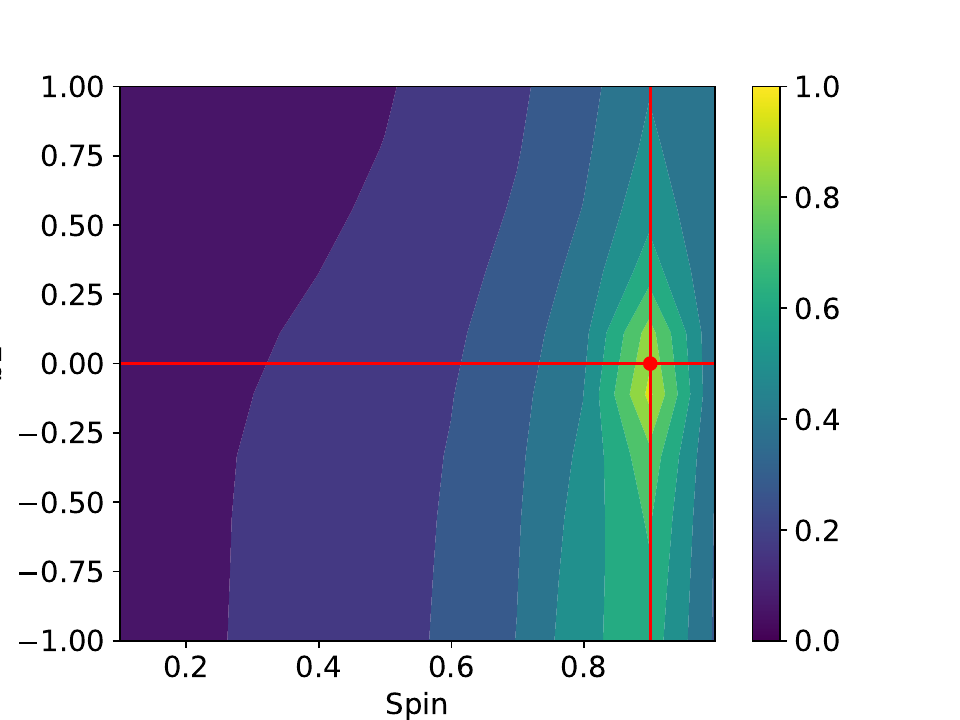}
    \caption{2-dimensional grid of the likelihood for spin vs b2.}
    \label{fig:spin_b2}
\end{figure}

\begin{figure}
    \centering
    \includegraphics[width=0.99\linewidth]{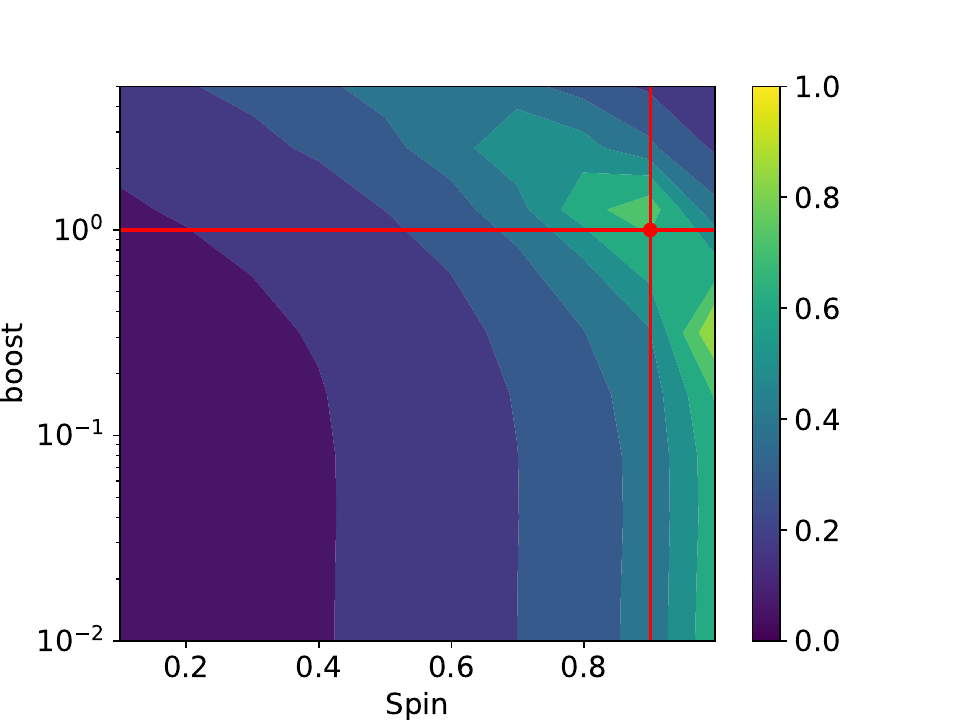}
    \caption{2-dimensional grid of the likelihood for spin vs boost.}
    \label{fig:spin_boost}
\end{figure}

\begin{figure}
    \centering
    \includegraphics[width=0.99\linewidth]{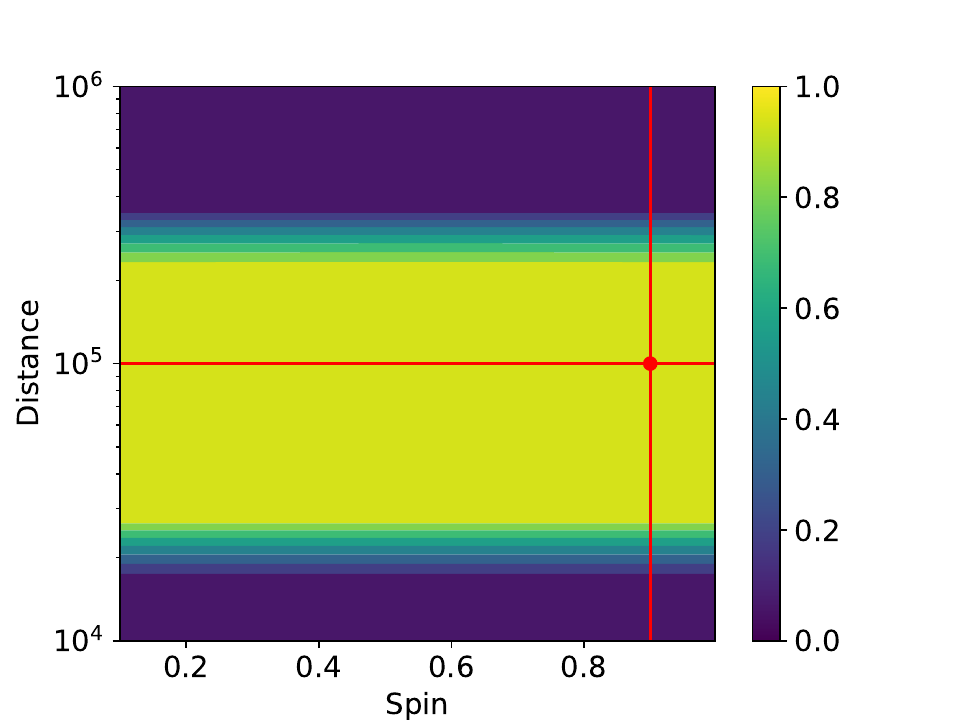}
    \caption{2-dimensional grid of the likelihood for spin vs distance.}
    \label{fig:spin_D}
\end{figure}

\begin{figure}
    \centering
    \includegraphics[width=0.99\linewidth]{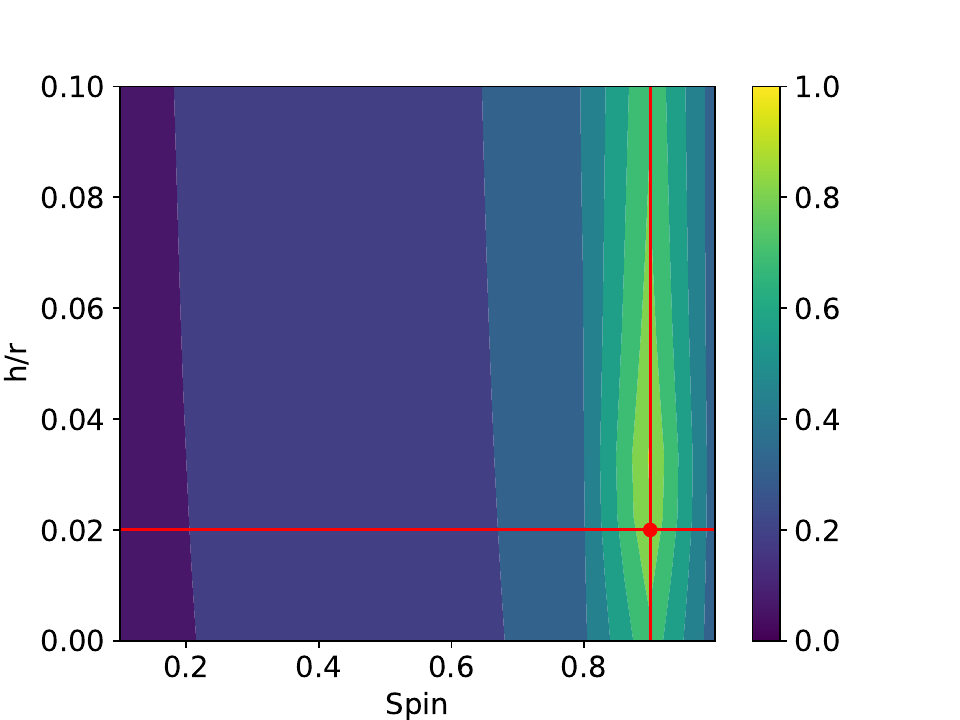}
    \caption{2-dimensional grid of the likelihood for spin vs scale height of the disk.}
    \label{fig:spin_honr}
\end{figure}

\begin{figure}
    \centering
    \includegraphics[width=0.99\linewidth]{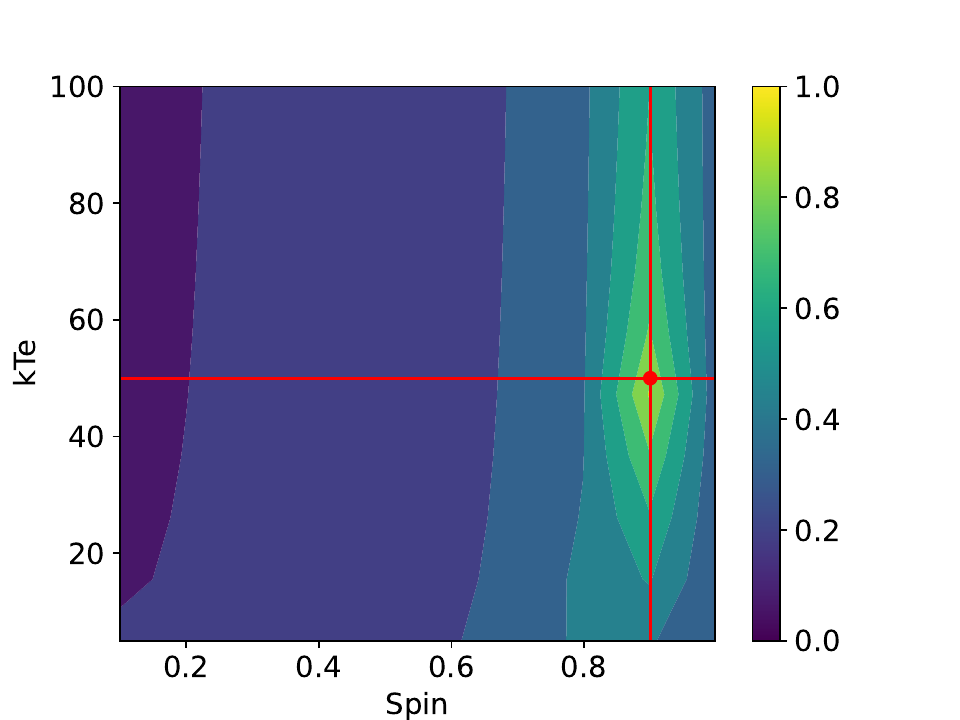}
    \caption{2-dimensional grid of the likelihood for spin vs kTe.}
    \label{fig:spin_kTe}
\end{figure}

\begin{figure}
    \centering
    \includegraphics[width=0.99\linewidth]{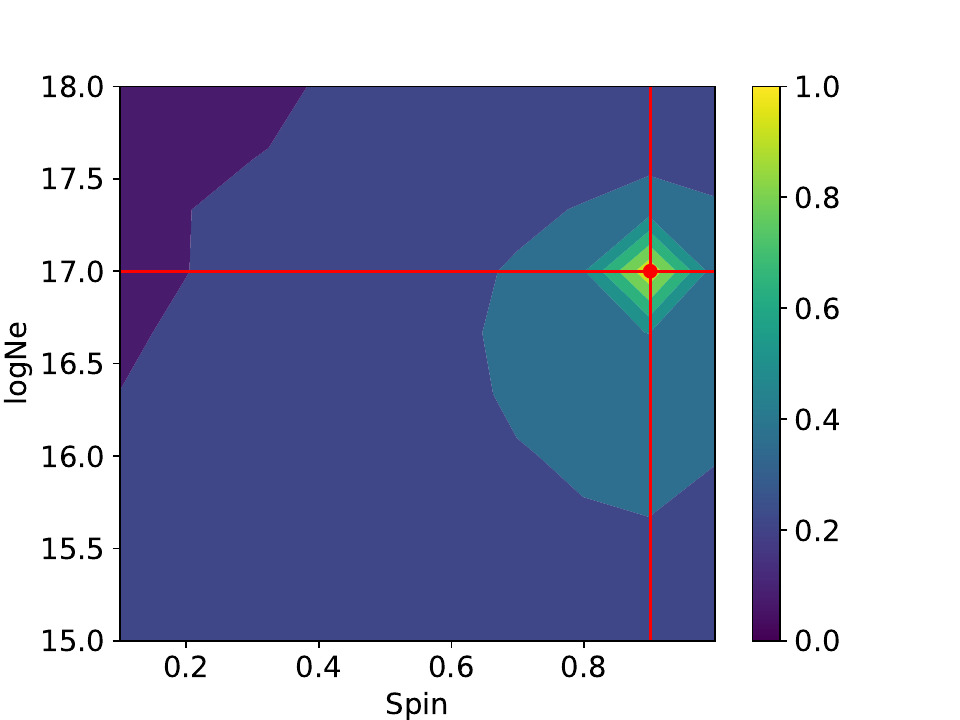}
    \caption{2-dimensional grid of the likelihood for spin vs $\log N_e$.}
    \label{fig:spin_logNe}
\end{figure}

\begin{figure}
    \centering
    \includegraphics[width=0.99\linewidth]{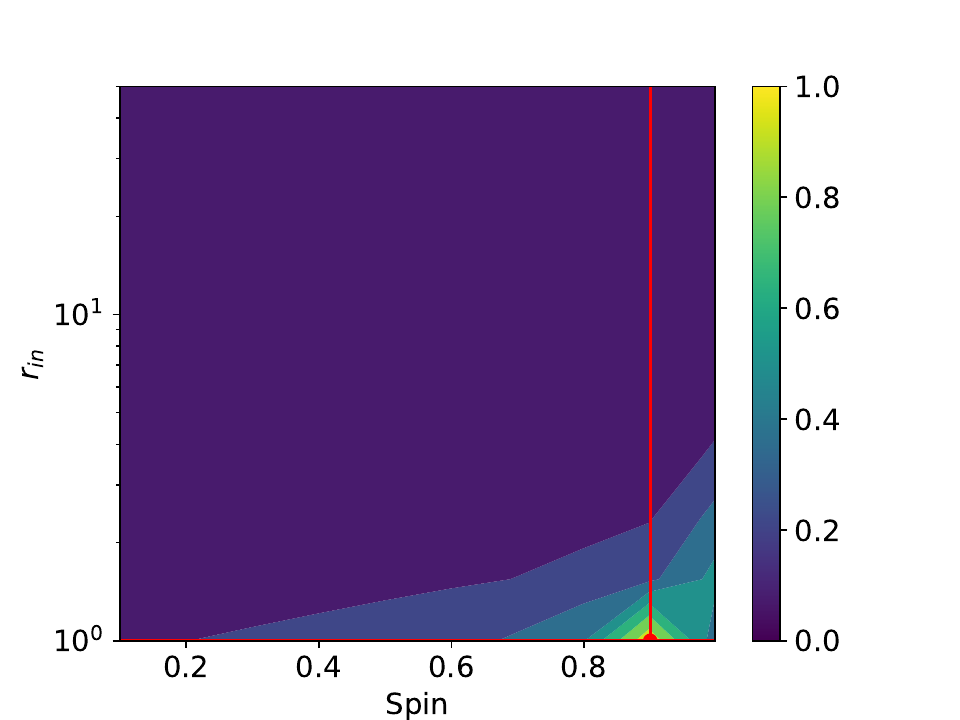}
    \caption{2-dimensional grid of the likelihood for spin vs inner radius.}
    \label{fig:spin_r_in}
\end{figure}

In this section, we show and further discuss the 2-dimensional biases for the inference in section \ref{fit}. In figures \ref{fig:spin_gam}and \ref{fig:spin_nh}, the maximum likelihood is very close to the true parameters when all other parameters are fixed. This supports that the emulator correctly captures the relationship between spin and $\Gamma$ and $N_h$.

Figures \ref{fig:h_spin} and \ref{fig:spin_i} show that the likelihood space is somewhat more complicated for inclination and height, which could potentially bias the inference to values of incorrect spin (with not as much an effect as Fig\ref{fig:2dlikelis}). Fig \ref{fig:h_spin} shows that, while the point of highest likelihood is near the truth, there are areas of higher likelihood at higher values of height where spin is lower. This is somewhat realistic (as increasing height allows for more photons to escape the influence of the black hole but decreasing spin also allows for more photons to escape the black hole due to less extreme general relativistic effects), but this likelihood space should be more clearly correlated. Fig \ref{fig:spin_i} shows a small split in the possible inferred inclination and spin of the system. The truth once again fits in the space of highest likelihood, and the correct degenerate relationship of high spin and low coronal height and vice versa resulting in similar reflection spectra due to relativistic effects is recovered. Figures C5-C15 show the remaining parameters correlated with spin, excluding outer radius due to it being very uninformative to fitting of the spectrum.

\begin{figure*}
    \centering
    \includegraphics[width=0.99\linewidth]{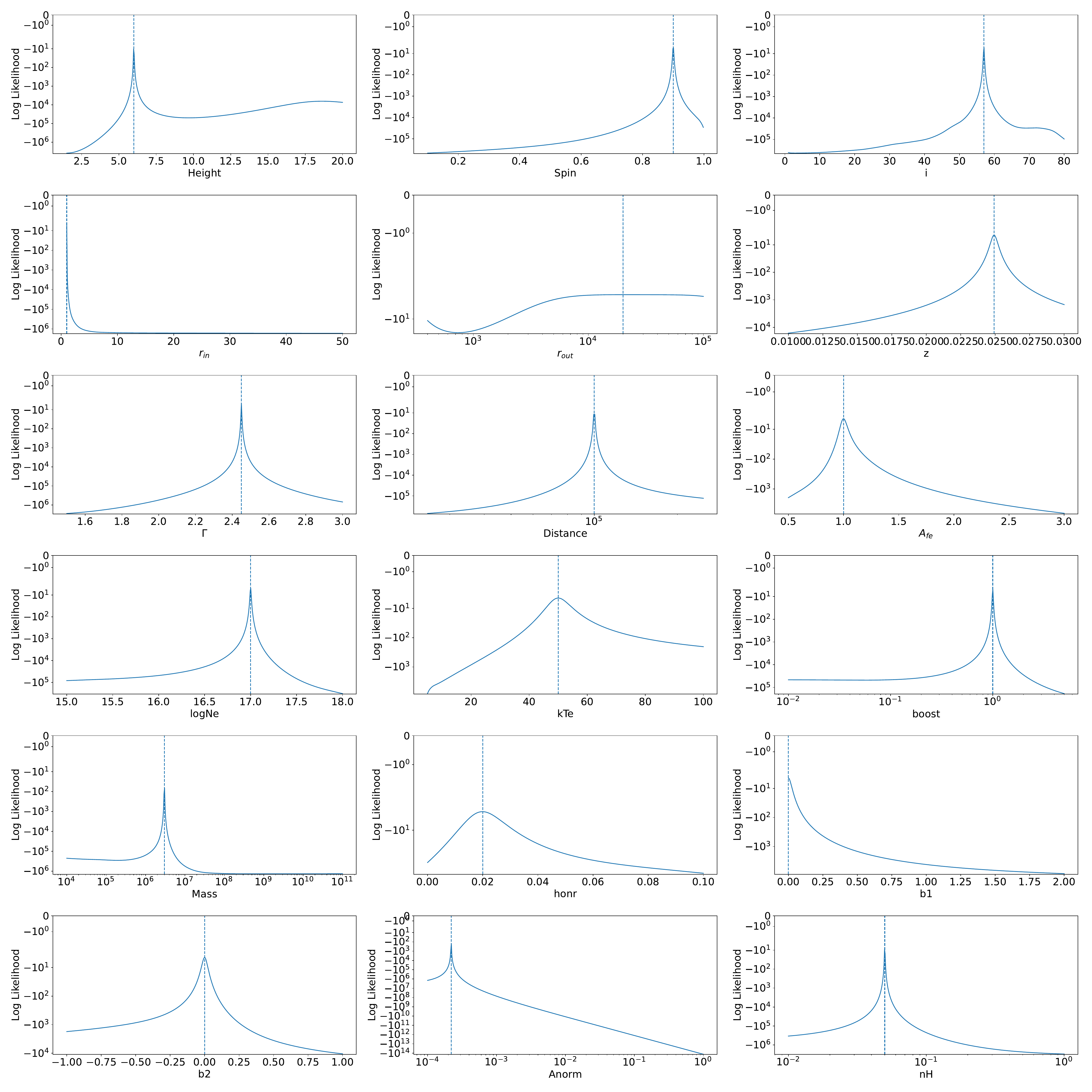}
    \caption{Plots of the log likelihoods as a function of each parameter while all parameters are fixed at the true value. The true parameter is plotted as a vertical dashed line.}
    \label{fig:likelihoods_1d}
\end{figure*}

Fig. \ref{fig:likelihoods_1d} shows the one dimensional likelihoods for each parameter. We observe that, when every other parameter is fixed, the emulator recovers the correct parameter value as the point of maximum likelihood. 

\section{Building emulators for physical models}\label{build_emus}

This section contains the insights we developed building RTFAST-spectra. These insights are intended to be at least somewhat spectral model neutral, though every problem in the machine learning context has its own quirks that prevent one-to-one translation between problems.

Feed-forward neural networks were overwhelmingly the best strategy in terms of emulation quickly and accurately. Other types of layers such as convolutions, and long short-term memory (LSTMs) were generally more inaccurate, or in the cases where they were of similar accuracy, slower than the methodologically simple linear network. Convolutional layers performed particularly poorly. We posit that, for many spectra, different emission lines in the spectrum are often too far apart within the spectrum to be encoded within the space of a single convolutional kernel. This means that important correlations are not captured by this type of neural network layer. A sufficiently wide enough kernel to capture these correlations quickly approaches that of a fully connected neural network, and loses any unique benefits offered by convolutional kernels.

Most notably, improvement in performance was overwhelmingly dominated by the addition of more data in training. We attempted several strategies in improving choice of data generation for training, but found that all more intelligent strategies involving attempted probing of the neural network's inaccuracies tended towards producing almost uniformly distributed points across the parameter space chosen once enough data had been generated. 

After a very large amount of data generated, we generally found that the neural networks improved in training logarithmically, with validation loss tightly linked to the training loss - not exhibiting evidence of over-fitting. While it is likely that we could have trained for much longer until true over-fitting had occurred, we found it more effective to build an ensemble to yield the improvement we desired. This also removed some systematic biases we found neural networks frequently exhibited when emulating spectra (this is particularly obvious in fig. \ref{fig:calib}). The ensemble generally mitigated systematic bias in the individual neural networks, and reduced the extremes of outlying bad performance. Adding more models into the ensemble seems to lead to a saturation in performance gains, but we could not test this extensively due to computational resource constraints. This generally implies that ensembles for these types of problems act much closer to the theoretical best case for ensembles. Ensembles work most effectively when errors in the neural network are from under-fitting rather than due to systematic bias in the training data. If the errors are approximately Gaussian due to the random initialisation of weights at the beginning, this means that simple averaging of the outputs of multiple neural network outputs will more closely approximate the truth.

Notably,the choice of loss functions also led to considerable improvements compared to the choice of model architectures. Choosing the standard Mean-Squared Error Loss  to train the network to emulate PCA components, we found the neural network to be almost unable to learn any relation between the parameters and PCA vectors. Weighting vectors allowed the neural network to first fit simple correlations and then learn the higher order correlations encoded in the PCA vectors as they became more relevant in the weighting during training. Some experiments we performed on non-decomposed forms of the original model were improved when we encoded the distribution of the entire spectrum into the loss via Kullback-Liebler divergence as an additional component to the mean-squared-error. As Kullback-Liebler divergence calculates the difference between the distributions, this encodes correlations between energy bins mathematically and punishes the neural network for not taking these correlations into account.


\bsp	
\label{lastpage}
\end{document}